\documentclass{IEEEtran}
\usepackage[utf8]{inputenc}
\usepackage[T1]{fontenc}
\usepackage{lmodern}
\usepackage{hyperref}
\usepackage[htt]{hyphenat}
\usepackage{amsmath}
\usepackage{multirow}
\usepackage{graphicx}
\usepackage{subcaption}
\usepackage{gensymb}
\usepackage{xcolor}
\usepackage{soul}
\usepackage{comment}
\usepackage{svn}  
\SVN $Author: damours $
\SVN  $Rev: 527 $
\SVN $Date: 2018-08-20 14:38:54 -0400 (lun 20 aoû 2018) $
\SVN  $URL: svn://localhost/greennetworks/trunk/elsarticle/article.tex $

	\title{Planning Solar  in Energy-managed Cellular Networks
		\thanks{Submitted to IEEE Access, August
			2018}} 
	\author{Mathieu D'Amours, Andr\'e Girard and Brunilde
          Sans\`o\thanks{M. D'Amours and B. Sans\`o  are with Ecole
            Polytechnique Montreal,
          A. Girard  is with INRS-EMT and GERAD,
            Montreal, Canada.}
	}

\begin{document}
\allowdisplaybreaks
\maketitle
\begin{abstract}
There has been a lot of interest recently on the energy efficiency and environmental impact of wireless networks.
Given that the base stations  are the network elements that use
most of this energy, 
much research has dealt with ways to reduce the energy used by the
base stations by turning them off during periods of low load.

In addition to this, installing a solar harvesting system composed of
solar panels, batteries, 
charge controllers and inverters is another way
to further reduce the network environmental impact and some research 
has been dealing with this for individual base stations.
 
In this paper, we show that both techniques are tightly coupled.
We propose a mathematical model that captures the synergy between
solar installation over a network and the dynamic operation 
of energy-managed  base stations. 
We study the interactions between the two methods for networks of
hundreds of base stations and show that the order 
in which each method is introduced into the system does make a
difference in terms of cost and performance. We also show that
installing solar is not always the best solution even when the unit
cost of the solar energy is smaller than the grid cost.
We conclude that planning the solar installation and energy management
of the base stations have to be done jointly. 
\end{abstract}

\section{Introduction}
\label{sec:intro}
There is a growing awareness to the fact that the communication
sector uses a significant amount of energy. This is especially true for
wireless, and in particular for the
base stations of the cellular networks, where energy costs make up a
large part of the operating expenses of service providers.

Two large trends appear in the literature to tackle the wireless
energy efficiency problem. 
On the one hand, there is the search for more energy-efficient transmission 
devices and technologies~\cite{auer12}.
This is mostly done  in  the physical layer and is thus outside the
scope of this paper since we focus here  on the network layer.
On the other hand,
new technology can be used
to improve energy consumption in the base stations, such as
\emph{sleep mode},  where some base stations
are turned off during low traffic periods and users 
can be re-allocated to the  active ones.

More recently, the idea of using green energy sources to power base stations
has been considered and
in particular, the use of solar energy~\cite{alsharif17} either as a
stand-alone source,
where the climate permits, or as an addition to conventional grid
electricity.
In most of the work on green energy, however, 
there is a common assumption that  green energy is either practically free or
at least much cheaper than grid power. From this assumption  follows the notion
that one should install solar equipment everywhere and that one should
use as much solar energy as possible.
This of course neglects the importance of the capital cost of the
solar equipment needed to  
implement energy harvesting.
In fact, there is a significant
cost to the purchase and  installation of solar collection equipment on base
stations which might lead to different solutions.

The literature has been missing not only a realistic view on solar, but, more importantly, 
 an integrated view of the networking and
solar energy issues that are important for a cost-effective green
energy network  planning. 
The objective of this paper is to challenge the fact that one can
manage the network via sleep mode
or provision for solar energy in an independent way.
In fact, we show that there is a close relationship between 
solar planning and  sleep mode.
Moreover, we find that the \emph{order} in which the different features
are optimized produce a very different outcome. We also find that
installing solar equipment everywhere is not always optimal
even when the unit cost of solar energy is smaller than the grid cost
Finally, we make the case for a joint network management-solar
planning optimization solution. 

\section{Previous Work}
\label{sec:prev}
\subsection{Base Station Energy Management}
\label{sec:relwork-bs-sleep}
It has been known~\cite{wu15,budzisz14,piovesan18} for quite some time that
turning off  base stations
during low traffic demand can yield significant energy savings. This
is possible if there are mechanisms for reallocating users to other
active base stations during that period, which is the case for current
technology. 
Therefore, a large body of literature in energy efficiency has been devoted to
modelling base station sleep mode. 

Some work~\cite{bousia12,bousia12a,bousia16,bousia16a,jia15} uses game
theory to compute the base station sleep schedule such that the total
power used is minimized subject to constraints on the quality of service
received by the users. A similar problem is studied in~\cite{cai13} in
the context of a heterogeneous network made up of macro and pico
cells and is extended in~\cite{ghazzai17} to the case of femtocells.
For 3G networks, the
minimum network energy 
use is computed in~\cite{wu13} 
where a base station is turned off whenever \emph{all} of its users can
be served by another active base station. This is based on a
time-independent Poisson distribution of users
where only the location of users enters into the calculation but where
the daily variation in demand is not taken into account.
The authors then use these results to
study the impact of the coverage radius and the cell size on energy
use.

Another energy-management technique is discussed in~\cite{le12a,chung15,ismaiil14} where base
stations can set their transmission power at different values.
This is also examined in~\cite{cili12,he14}
where coordinated multipoint transmission is used 
to reach isolated users  instead of increasing the base station power.
Another cooperative solution is
studied in~\cite{oikonomakou17} using game theory.

Concerning the network dynamics and savings, 
there has been a number of  models  to optimize the operation of base
stations in 3G networks~\cite{wu15}. Most of these are real-time
algorithms where a quasi-optimal policy is computed given an already
existing network. Although these can provide substantial energy
savings, they are constrained by the structure of the network in which
they operate.
A similar model is discussed in~\cite{classen13} where the objective
of reducing energy use 
is to minimize two objective functions: the number of installed based
stations and the number of users not served by any base station. 

The idea that sleep mode management must be integrated at the planning stages 
was formalized in~\cite{boiardi12,boiardi13,boiardi14} where it was shown that
the joint optimization can bring savings of up to 30\% with respect to individual optimization. 
\subsection{Solar-Powered Base Stations}
\label{sec:relwork-solar}
The issue of using solar energy to power base stations has
received some attention over the last few years. This has been
examined first for base stations in isolated areas where grid power
is either not available or it is very expensive.
An example of a  simple case is~\cite{marsan13} that studies whether a
base station can be 
powered exclusively with solar energy taking into account
the traffic load and sleep mode. A
statement is made to the effect that capital and operating costs are
such that a purely solar power base station is competitive with diesel
generation. 

The most recent work on using solar power for 3G has been
investigated in~\cite{alsharif17}. The objective is to choose the
solar equipment of an UMTS Node B to minimize their net present
cost. The hourly load is available as well as the average monthly
solar power. The solution method is to compute the 
power available at each hour for all possible equipment configuration
and choose the best one. 
The energy needed to serve traffic is not taken into account. The only
requirement is that the total power available from the batteries and
panels should equal the power needed by the base station plus
losses. The main
conclusion is that solar energy is a realistic
option even for 3G technology.

Another aspect of solar energy optimization is to take into account the
random variations in solar energy. 
Stochastic programming was used in~\cite{niyato12}  to optimize
the expected cost of purchasing energy from the grid
under uncertainty due to solar energy availability, variable traffic load
and on-demand grid prices. The decision variables
were the amounts of electricity to buy from the grid during the day.

The work of~\cite{rubio14}  models the on/off switching strategy for a
base station operating only on solar energy. It considers the
case of two base stations and proposes a solution based on a robust
Bayesian technique assuming perfect
information on the traffic.

Network effects can also arise due to the presence of a power
distribution smart grid.  The energy
management for a given base station  
connected to a smart grid is studied
in~\cite{leithon14a,leithon14b,farooq17}. 
A critical upper
bound on the batteries' capacity where no more energy savings can be
obtained is investigated in~\cite{leithon14a}.  The charge of the
batteries is also studied in~\cite{mendil16}, where the charge needs to
stay within a given range. They model the charge and discharge to optimize
the life expectancy of the batteries.  
Note that in the present paper, we do not optimize battery capacity
which is  sized through preliminary tests.

A hybrid solar-grid system for a single base station that is less
costly than pure solar is 
proposed in~\cite{zhang15b}. An important feature of this work is that
the objective is to  minimize the total cost made up of the capital cost of
panels and batteries  plus the operating cost of using grid power. 
The number of batteries and the size of the solar panels are optimized
as well as the energy management.
This seems to be one of the rare cases where planning is done for an
horizon of many years but not in a network context. This is
solved in a two-step process. First, a 
model is set up for the optimization of the base station in a single
year. The multi-year problem is solved by a sequence of single-year
optimizations.
\subsection{Solar and Sleep Mode}
\label{sec:solarandsleep}
The decision to use solar energy or not can be made for each base
station separately, based on cost, performance, etc. Such decision
has then no impact on other base stations. Things get much more
complicated when \emph{both} solar power and sleep
mode are used. The reason is that the decision to put a base station in sleep
mode means that the users have to be reallocated to other base
stations, thus increasing their load. This in turn may impact the
decision in these other base stations to use solar energy during that
time, for instance because the solar panels cannot provide enough
energy to serve all the users. In other words, the two problems become
tightly coupled and must be solved together. Because of this
difficulty, there has not been much work done on this topic.

We have found only one reference~\cite{wang17} where both solar and
sleep mode are used together. The objective is to minimize the total
energy cost, either solar or grid, where decision variables are used
to determine 
the assignment of users to base stations and the use of grid or solar
power. There is the usual assumption that ``the unit cost of green
energy is cheaper than that of the on-grid energy'' and the authors
also assume batteries with infinite capacity. The solution technique
based on  a long-term demand
forecast is used to compute a target sleep schedule. The main
differences with our approach is that the model has no limit on
battery storage, the costs are not based on real estimates since only
the ratio of solar to grid energy is used and all base stations are
assumed to have solar energy available. 
\subsection{Costs}
\label{sec:costs}
Much of the work on solar energy assumes that its cost is very small
compared with the grid cost and it is very seldom taken into
consideration. One exception is the work of~\cite{han16} that
optimizes the sizing 
of the solar panels installed
on macro base stations and the  battery
banks capacity to minimize the capital cost. The trade-off is between
installing the solar systems on macro base stations on the one hand
and off-loading traffic on small base stations on the other.
The fraction of the
total energy requirement that is served by the green sources is given
a priori and is not a result of the optimization, which is an
important difference with what we do. Also, there is no actual value
for the costs and the results are presented as a function of the cost
ratio between the two types of energy.

It is also worth mentioning that  there is little work trying  to estimate realistic
costs for the whole solar 
energy harvesting system. Most
papers consider solar panels and batteries but leave out charge
controllers and inverters which are an 
important part of the equipment. In fact,  inverters and
charge controllers not only increase the
capital cost of the system but also decrease its efficiency.  

In the literature we reviewed, no work considered inverters and charge controllers,
with the exception of~\cite{alsharif17} that takes into account the cost, lifetime and
efficiency of the inverters, but not  the charge controllers.

Summarizing, 
we are not aware of any research that 
integrates solar panels, batteries, inverters and
charge controllers with the sleep-mode of the equipment into a
long-range planning model. 
\section{System Model}
\label{sec:model}

The most important feature that sets our model apart from
the previous work  is that it is a
\emph{planning} model where we choose if and where  to install solar
equipment.
This equipment will be used for a long time, typically many years, so that
the model has to be based on a long-term view of the network,
typically 10 years or more.

The actual network
operation, on the other hand,  depends on some time-varying
quantities, like user demand or solar illumination, that are generally
random and  vary
on a much shorter time scale, typically hour or even 
shorter.  In general, it is not possible to model these short-term
features in detail over the whole planning horizon because of memory
or computation
time limitations. In this paper, they are replaced
by a long-term estimate of the performance of the short-term
operation, generally measured from historical data, that
corresponds to some important time, say the busiest day of the
year. Typical performance values could be for instance the average hourly solar
illumination averaged over the horizon, or the average traffic
demand over that period.
The estimation of these long-term approximations is a difficult
problem by itself and is outside the scope of this paper, where we
simply assume that they are known.

In addition, our model integrates some features of a number of
previous work into a coherent mathematical formulation. It also has
the following features that are 
different from the state of the art.

First, as mentioned above, this is a long-term planning model over a
horizon of many years, used to decide if and where solar equipment
should be installed. 
This is traded off against the operating cost
of using grid power.
In order to get realistic results,
we explicitly take into account the capital cost needed to
implement solar power using cost values currently available. As we
mentioned before, adding inverters and 
charge controllers can change significantly the cost of the solar
energy.
We also model the batteries costs and limitations, something that is
not always done in some models. 

The second important difference is that  we 
model a whole network, where it is
possible to off-load traffic from one base station to another. 
Also, we introduce the notion of an equivalent day used to
model to some extent the impact of the daily traffic and light
variation on the long-term planning decisions.

Finally, we formulate the model as a linear integer
program so that it can be computed by standard solvers for small
enough networks. This allows us to gain insight into the more
important factors that affect the decision to implement solar energy
or not. 

Our model is based on~\cite{boiardi13} where 
the network is made up of a given set of base stations that can turn
their transmitter on or off during the day in order to save energy.  Traffic is
generated by so-called \emph{traffic test points},
devices that aggregate traffic from users. 
The day is divided into a number of time periods that don't have to
have the same length, e.g., day or night, or hourly during the day and
a single night period. At the beginning of each time period, a
decision has to be made for  each base station to switch it on or off
and how to allocate all test points in such a way that they all fall
within the coverage region of at least one base station that is turned on. 

We assume that the base stations, traffic and
connection test points are given.
The model of~\cite{boiardi13} is  extended by having the possibility
of installing solar 
cells and batteries to power the base stations. 
The solar cells feed
into a battery pack which can then  be used to feed the base station instead
of using grid power. 
The decision to install solar cells is taken
once for each base station
at the beginning of the study period.

The base stations
can be in two states, idle or active. In the idle state, the antennas
are turned off and the base station uses minimum power. In that
state, the base 
station cannot serve any user. In the active state, the
antennas require an extra amount of power to serve the users.
At the beginning of each time period, each base station must choose
the state, idle or active, and whether the corresponding energy comes
from the grid or the batteries.
\section{Mathematical Formulation}
\label{sec:math}
\subsection{Sets}
\label{sec:set}
First we define the following sets:
\begin{description}
\item[$S$] Installed base stations
\item[$I$] Test points
\item[$T$]  Time periods. 
\end{description}
The time periods are indexed by 
$t = 1 \ldots T$. This is called the \emph{time base} in the following
and each value indicates when the period starts, e.g., a time base $T
= \{0,8,16 \}$ is made up of 3 periods starting at 0:00, 08:00 and
16:00 hours.

In general, $j$ refers to a base station, $i$ to a test point and $t$
to a time period. These indices are always assumed to run over their
whole set $S$, $I$ or $T$ unless otherwise noted.
\subsection{Parameters}
\label{sec:paramdef}
These are known network parameters. In some cases, they  readily
available and if not, they can be
calculated with realistic data as explained in Section~\ref{sec:param_size}.

The objective
function is the sum of capital costs and the total value of the
operating costs over all days of the study period with
\begin{description}
\item[$\phi$] The number of days over the time horizon for
the planning, e.g., for a planning horizon of $M$ years, 
$ \phi = 365\, M$
\item[{$C_j^S$}] Installation cost for solar panels, 
  batteries, inverters and charge controllers on base station $j$. The cost includes capital as well as
  replacement cost and can depend
  on the particular base station and type of solar module.
\item[$C^E_{j}$] Grid energy cost. We assume  that this is constant
  over the length of the study.
\item[$E^S_{j,t}$] The amount of electrical energy produced by the
  solar panels at base station $j$ during interval $\Delta_t$.
\item[$E^0_{j}$] Energy needed by the base station in the idle state
\item[$E^1_{j}$] Energy needed by the base station in
  the active state.
\item[$E^T_{i,t}$] Energy required by test point $i$ in period
  $t$.
  \item[$N^{tp}_{bs}$] The number of test points per base stations 
\item[$B_j^+$] Maximum battery capacity. This is the total  energy
  that can be stored by 
  \emph{all} batteries installed in a base station.
\item[$B_j^-$] Minimum battery capacity. This is
  some fraction of $B^+_j$.
\item[$k_{i,j}$]  Indicator function set to 1 if traffic test point 
$i \in I $ 
is covered by the base station installed in $j$ and 0 otherwise.
\end{description}
Based on these parameters, 
we describe the linear model: parameters, variables and constraints
corresponding to the base stations and test points.
\subsection{Variables and Constraints}
\label{sec:base-station}
These are the optimization variables that correspond directly to the
operation model. They are set to~1 if
\begin{description}
\item[$z_j$] Solar equipment is installed at base
  station $j$
\item[$x^o_{j,t}$] The base station is in the idle state in interval $t$
\item[$x^b_{j,t}$] The base station uses battery power during
  interval $t$
\item[$h_{i,j,t}$] Test point $i$ is assigned to base
  station $j$ in period $t$
\end{description}
and~0 otherwise.
We also need some intermediate variables to simplify the presentation.
\begin{description}
\item[$D_{j,t}$] Energy required by the users assigned to $j$ in $t$
\item[$E^P_{j,t}$] Energy used by the antennas of $j$ in $t$, whether
  it is in the idle or active state
\item[$L_{j,t}$] Energy lost because of the limited capacity of the
  batteries in base station $j$ in period $t$
\item[$E^B_{j,t}$] The amount of energy available in batteries at base
  station $j$ at the beginning of interval $t$.
\end{description}
These variables are subject to constraints. First we cannot use
battery power unless solar equipment has been installed:
\begin{align}
  x^b_{j,t} \le z_j \quad \forall j \in S, \ \forall t \in T. \label{eq:bndxbz}
\end{align}
Next,  the user
demands are computed from the assignment variables
\begin{align}
  D_{j,t} = \sum_{i\in I} E^T_{i,t} h_{i,j,t} \quad \forall j \in S, \ \forall t \in T  \label{eq:lindemand}
\end{align}
where the assignment variables are subject to the constraints
\begin{align}
  \sum_{j\in S} h_{i,j,t}   = 1 \quad \forall i \in I, \ \forall t \in T \label{eq:linuniqueh}.
\end{align}
We also assign the test points only to base stations that are not in idle state:
\begin{align}
  x^o_{j,t} \le 1 - h_{i,j,t} \quad \forall i \in I, \ \forall j \in S, \ \forall t \in T.  \label{eq:notoff}
\end{align}

Note that there are no explicit coverage constraints in the model. These can be taken into account in a number of ways. The easiest one would be to add constraints of the form
$h_{i,j,t} \le k_{i,j}$ which would prevent the allocation of test
points to base stations when they are not within the coverage. Most
modern solvers would recognize this condition in the pre-solve phase
and automatically remove the variables. Another way would be to define
the $h$ variables only for those cases when the test point is within the
coverage radius. Finally, one can simply fix the $h_{i,j,t} = 0$ in the AMPL
script whenever $k_{i,j} = 0$. This is the solution we take here.

The next set of constraints describes the energy production and
management. First, we can compute the energy used by the antennas as
\begin{align}
  E^P_{j,t} = E^0_{j,t} x^o_{j,t} + E^1_{j,t} ( 1 - x^o_{j,t}) \quad
\forall j \in S, \ \forall t \in T . \label{eq:defep}
\end{align}
The  demand must not exceed the energy
available to the antennas in the idle or active state, which yields
\begin{align}
   D_{j,t} \le E^P_{j,t} - E^0_{j,t} \quad \forall j \in S,  \forall t \in T.   \label{eq:transconst}
\end{align}
Replacing~\eqref{eq:lindemand} and~\eqref{eq:defep}
in~\eqref{eq:transconst}, we get
\begin{equation}
  \sum_{i\in I} E^T_{i,t} h_{i,j,t}  \le (E^1_{j,t} - E^0_{j,t}) ( 1 - x^o_{j,t}) \quad \forall j \in S, \ \forall t \in T
  . \label{eq:demandbnd}
\end{equation}
Next, we must model the assumption  that we can
use only one of the two energy sources, solar or grid, during a given
period and that  the decision whether  to use
solar or not is made at the beginning of each period only. This means
that we can use solar energy only if the amount of energy stored in
the battery at the beginning of the period plus that produced by
the solar panels is at least as large as the required energy during
that period without depleting the batteries beyond their minimal
value.

In order to simplify the notation, we define $\overline{E}^P_{j,t}$,  the antenna
power used in battery mode, as
\begin{align}
\overline{E}^P_{j,t} =   x^b_{j,t} E^P_{j,t} && \forall j
\in S, \ \forall t \in T . \label{eq:defebar}
\end{align}
The constraint on energy use can then be written
\begin{align}
  E^B_{j,t} + E^S_{j,t} -  \overline{E}^P_{j,t} & \ge B^- && \forall j
\in S, \ \forall t \in T . \label{eq:batteryfeed}
\end{align}
We also have to express the fact that the excess energy remaining
at the end of a period is stored in the battery and is available at
the beginning of the next one. If we impose the condition that all the
energy must be stored, we would have a constraint of the form
\begin{displaymath}
  E^B_{j,t}  = E^B_{j, t-1}  + E^S_{j,t-1}  - \overline{E}^P_{j,t-1}.
\end{displaymath}
If the value on the right-hand side turns out to be larger than $B^+$, the
constraint can be satisfied only by not using solar during that
period, which is not a realistic solution. Instead, we write the
condition as an inequality constraint
\begin{displaymath}
  E^B_{j,t}  \le E^B_{j, t-1}  + E^S_{j,t-1}  - \overline{E}^P_{j,t-1}
\end{displaymath}
so that the excess energy can be used up to the value of $B^+$ but no
more. We also define the slack variables $L_{jt}$ explicitly since
they represent the lost energy in that period. We then have
the amount of energy stored in the batteries at the beginning of time
$t$ as
\begin{align}
E^B_{j,t} & =  E^B_{j, t-1}  + E^S_{j,t-1} - L_{j,t-1} -
  \overline{E}^P_{j,t-1}, \label{eq:dynamics}  \\
  B^-_j & \le E^B_{j,t} \le B_j^+  \quad \forall j \in S,  t \in T , \label{eq:linchargeub} \\
  0 & \le L_{j,t} \le E^S_{j,t} \quad \forall j \in S, \ \forall t \in T . \label{eq:bndl}
\end{align}
Constraint~\eqref{eq:bndl} is due to the fact that  the maximum amount
of energy we can lose is that produced by the solar panels.

Finally, we need to impose 
a condition related to the assumption that the energy
use during a single day is in a sense
representative of the operation of the network over the time
horizon. This means that this pattern will repeat
itself every day. In that case, the excess energy at the end of the
last period is the available energy at the beginning of the next
day. For consistency, we impose the condition that
\begin{align}
  E^B_{j,1} = E^B_{j,T} + E^S_{j,T} - L_{j,T} - \overline{E}^P_{j,T}
  \quad \forall j .
  \in S \label{eq:cycle}
\end{align}

The objective function is the sum of the capital cost and the grid cost
\begin{align}
  C & = \sum_{j\in S} C^S_{j} z_j + \phi \sum_{\substack{j\in S \\ t\in T}} C^E_{j} E^P_{j,t} ( 1 -
      x^b_{j,t} ) . \label{eq:totcost}
\end{align}
As written, the problem is not linear since it contains quadratic
terms of the form $x^o_{j,t} x^b_{j,t}$ coming from~\eqref{eq:defep}
and~\eqref{eq:defebar}. We can linearize it by
defining supplementary variables and constraints
\begin{align}
  w_{j,t} & = x^b_{j,t} x^o_{j,t} \label{eq:defw} \quad \forall j \in S, \ \forall t \in T, \\
  w_{j,t} & \le x^b_{j,t} \label{eq:conswxo} \quad \forall j \in S, \ \forall t \in T, \\
  w_{j,t} & \le x^o_{j,t} \label{eq:conswxb} \quad \forall j \in S, \ \forall t \in T, \\
  w_{j,t} & \ge x^b_{j,t} + x^o_{j,t} -1 \quad \forall j \in S, \ \forall t \in T. \label{eq:conswxoxb}
\end{align}
The quadratic term $E^P_{j,t} x^b_{j,t} $, which represents the antenna
power used in battery mode,  appears
in equations~\eqref{eq:batteryfeed},~\eqref{eq:dynamics}
and~\eqref{eq:totcost}. It can be written  in terms of $w$ as
\begin{displaymath}
  x^b_{j,t} E^P_{j,t} = E^0_{j,t} w_{j,t} + E^1_{j,t} \left(x^b_{j,t} - w_{j,t} \right)
\end{displaymath}
which is now a linear term.
\subsection{Linear Optimization Model}
\label{sec:line-optim-model}
We now summarize the complete linear model where the equation labels
refer to the original definitions. Recall that unless otherwise noted,
the indices 
run over the complete sets $  i \in I $, $t \in T$   and $ j \in
S$. In the  following, we denote this as \emph{problem PL}.

\begin{align*}
 \min_{z,x,h,w} C & =  \sum_{j\in S} C^S_{j} z_j +\nonumber \\
  & \sum_{\substack{j\in S \\ t\in T}}  C^E_{j,t} \left[
  \right.
                    E^0_{j,t} x^o_{j,t} + E^1_{j,t} ( 1 - x^o_{j,t} ) \nonumber \\
& \left.  - E^1_{j,t} x^b_{j,t} + w_{j,t} \left( E^1_{j,t} - E^0_{j,t}
  \right) \right]
  \tag{\ref{eq:totcost}}
\end{align*}
\begin{align*}
 x^b_{j,t} & \le z_j \tag{\ref{eq:bndxbz}}    \   \\
 \sum_{j\in S} h_{i,j,t} & =  1 \tag{\ref{eq:linuniqueh}}    \   \\
  x^o_{j,t} & \le 1 - h_{i,j,t} \tag{\ref{eq:notoff}}   , \   \   \\
E^B_{j,t} & + E^S_{j,t} - E^0_{j,t} w_{j,t} - \nonumber \\
&  E^1_{j,t} \left( x^b_{j,t}  - w_{j,t} \right)  \ge B^- \tag{\ref{eq:batteryfeed}}\\
E^B_{j,1} & = E^B_{j,T} + E^S_{j,T} - L_{j,T} \nonumber \\
&  - E^0_{j,T} w_{j,T} - E^1_{j,T} \left( 
x^b_{j,T}  - w_{j,T} \right) &  \tag{\ref{eq:cycle}} \\
   E^B_{j,t} &= E^B_{j, t-1}  + E^S_{j,t-1} \nonumber \\
& - L_{j,t-1}  - E^0_{j,t-1}
               w_{j,t-1}\nonumber \\
 & - E^1_{j,t-1} \left( x^b_{j,t-1} - w_{j,t-1} \right)
               \tag{\ref{eq:dynamics}} \\
   \sum_{i\in I} E^T_{i,t} h_{i,j,t} &  \le (E^1_{j,t} - E^0_{j,t}) ( 1 - x^o_{j,t})
 )    \   \tag{\ref{eq:demandbnd}}\\
 B^-_j & \le E^B_{j,t}   \le B_j^+    \   \tag{\ref{eq:linchargeub}} \\
 0 & \le L_{j,t} \le E^S_{j,t}    \    \tag{\ref{eq:bndl}} \\
 w_{j,t} & \le x^b_{j,t}    \   \tag{\ref{eq:conswxo}} \\
 w_{j,t} & \le x^o_{j,t}    \   \tag{\ref{eq:conswxb}} \\
 w_{j,t} & \ge x^b_{j,t} + x^o_{j,t} -1    \   \tag{\ref{eq:conswxoxb}} 
\end{align*}

\section{Sizing the Parameters}
\label{sec:param_size}
\begin{figure}
	\centering
	\includegraphics[scale=0.55]{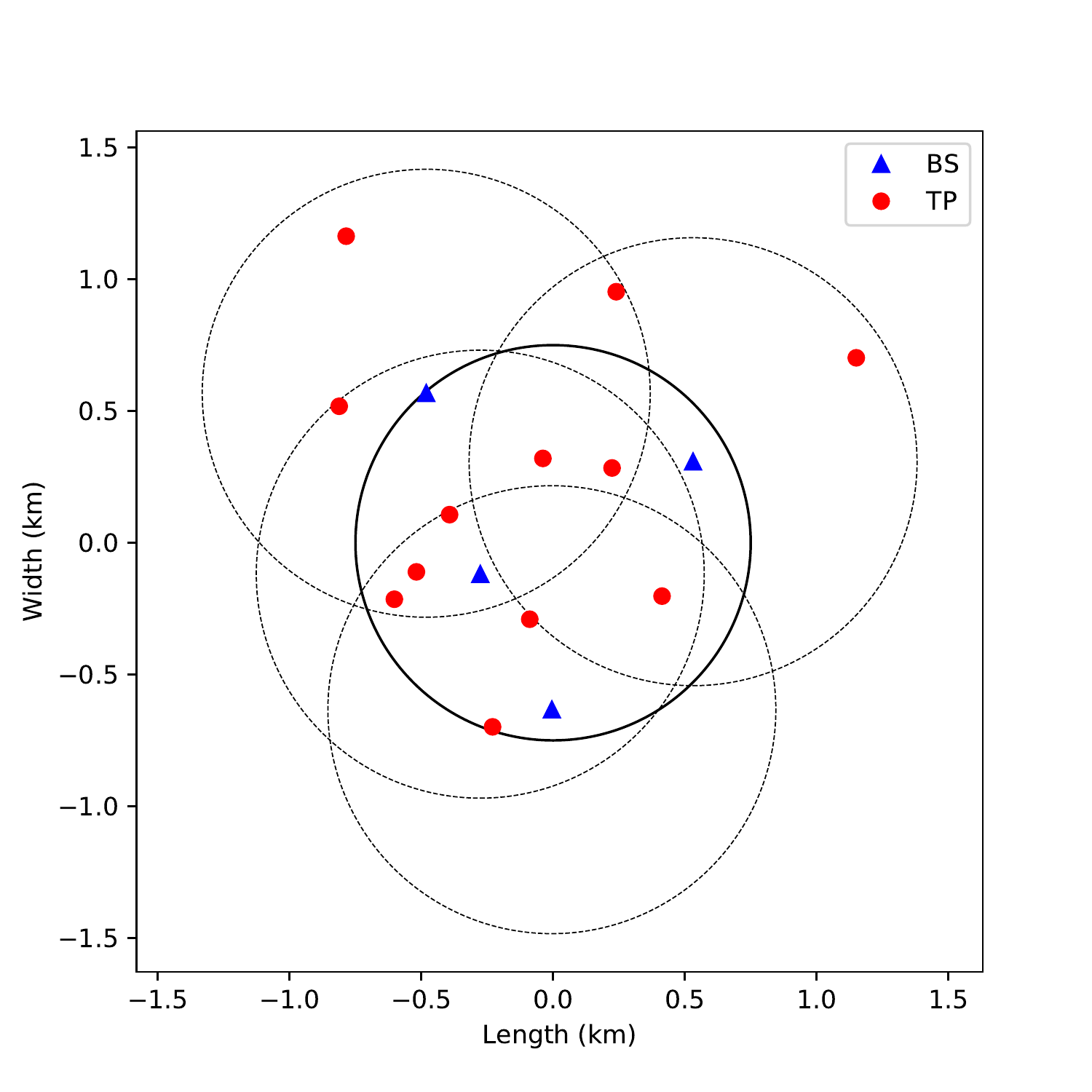}
	\caption{Small Network}
	\label{fig:net}
\end{figure}

In this section we
discuss how to size the parameters of Section~\ref{sec:math}:
the grid and solar equipment costs, the number
of solar panels, the
batteries that have to be installed to power a micro base station  and
the time intervals used to approximate the solar and usage profiles.
We also show in Fiture~\ref{fig:net} the small network that will be
used later on for explaining some of our results.
\subsection{Equipment Cost}
\label{sec:equipcost}
We model four types of solar equipment:
the solar panels, denoted by $S$, charge controllers, by $C$,
inverters, by $I$  and batteries, by $B$. These should not be confused
with the set notation used elsewhere in the paper.
Each
equipment type $K \in \{S, C, I, B\}$ has a number of specific
parameters and the following common parameters: 
\begin{description}
\item[$N^K$] The number of equipment units installed in the base
  station
\item[$L^K$] The  lifetime of each unit
\item[$C^K$] Cost of one unit
\item[$R^K$] The number of times the equipment has to be replaced
  during the study period
\end{description}
The number of replacements is given by
\begin{equation}
  R^{K} = \left \lceil{\frac{\phi}{L^K}}\right \rceil ,
  \label{eq:R_k}
\end{equation}
where $\phi$ is the length of the planning horizon.
The cost of solar equipment is then given by
\begin{equation}
  C^S_j = \sum_{K} C^K_j N^K_j R^K_j .  \label{eq:csj}
\end{equation}
The number of solar panels and batteries $N^S$ and $N^B$ is fixed by the user. 
$N^I$ and $N^C$ are then calculated so that the inverter power and the charge 
controller current match the solar panels power and batteries voltage. 

To obtain a realistic cost, the equipment parameters have to be
computed from 
real data. The ones used in this paper are all in the low price
range. Also, power cables and possibly other devices are not taken  
into consideration which is one of the reasons why this model is
somewhat biased in favor of  solar energy. 

Table~\ref{tab:equipparam} shows the different parameters with the values that
have been used in this paper. Note that the batteries \emph{DoD} is
the \emph{Depth of Discharge}, which is the limit the batteries charge
can be depleted. For more information about these equipment, the
interested reader can  find this information in~\cite{panelYangtze,battRolls,invJYins,chargeContrHanfong}.

\begin{table*}[]
	\centering
	\begin{tabular}{ccccccc}
		\hline
		\multicolumn{7}{|c|}{Solar Panel}\\
		\hline
		Cost  & Area  & Efficiency & Power &  & Degradation rate & Lifetime \\
 		(USD) & (m$^2$) & (\%) & (W) &  & per year (\%) & (years)  \\
 		112 & 1.62 & 18.03 & 280 &  & 0.5 & 20 \\
 		 &  &  &  &  &  &  \\
		\hline
		\multicolumn{7}{|c|}{Battery}\\
		\hline
		Cost & DoD & Efficiency & Voltage & Capacity & Degradation rate & Lifetime \\
		 (USD) & (\%) & (\%) & (V) &  (VAh) &  per year (\%) & (years)  \\
 		345 & 50 & 90 & 6 & 428 & 3 & 7 \\ 	
 		 &  &  &  &  &  &  \\
 		 
		\hline
		\multicolumn{7}{|c|}{Inverter}\\
		\hline
		Cost &  & Efficiency & Power &  &  & Lifetime \\
		 (USD) &  & (\%) & (W) &  &  & (years) \\
 		140 &  & 90 & 2000 &  &  & 10 \\ 	
 		 &  &  &  &  &  &  \\
 		 
		\hline
		\multicolumn{7}{|c|}{Charge Controller}\\
		\hline
		Cost &  & Efficiency & Current &  &  & Lifetime \\
		 (USD) &  & (\%) & (A) &  &  & (years) \\ 		
		26 &  & 95 & 60 &  &  & 10
	\end{tabular}
	\caption{Solar Equipment Parameters}
	\label{tab:equipparam}
\end{table*}
\subsection{Solar Energy}
\label{sec:solarenergy}
We compute $E^S_{j,t}$ from the electrical output power
$\overline{W}^S$ of a solar module that, at some instant 
$t$, is directly proportional to the solar radiation $G(t)$ at that
time~\cite{marsan13}. In particular, for the maximum value
$\overline{G}$, which depends on the region where the module is
installed, and a given module surface area $A$, we get
\begin{equation}
\overline{W}^S = \overline{G} A \eta^S . \label{eq:W^S_max}
\end{equation}
The  electrical power  $W^S_t$ produced in a given  period $t$ is given by
\begin{equation}
W^S_t = \overline{W}^S \pi^S_t ,  \label{eq:W^S}
\end{equation}
with $\pi^S_t$ as the fraction of maximal solar radiance during the
period of the day $t$ (See Section~\ref{sec:net-def} for a description
of $\pi^S_t$).

The total energy produced by a module during that period is
\begin{equation}
E^M_{j,t} = W^S_t \Delta_t . \label{eq:calc_E^M}
\end{equation}
if we install $N^S_j$ solar modules,  the total solar energy  they
produce can be computed 
using~\eqref{eq:W^S_max},~\eqref{eq:W^S} and~\eqref{eq:calc_E^M} and
is given by
\begin{align}
E^S_{j,t} & = N^S_j   E^M_{j,t} \nonumber \\
& = N^S_j \overline{G} A \pi^S_t \Delta_t \prod_k^K \eta^k \label{eq:calces}
\end{align}
where $\eta^k$ is the efficiency of each solar equipment $K \in \{S, C, I, B\}$.
\subsection{Test Point Energy}
\label{sec:tpenergy}
We assume that in the busiest time of the day with test points at
their maximum load, there is enough base stations to feed the 
test points. That means that, at that time,  
all of the energy available in the set of base stations is used. The
difference between the base stations  
energy in thee active and idle states is the energy needed to power
the test points.
\begin{equation}
E^T_{i,t} = \frac{1}{N^{tp}_{bs}} \left(E^1_j-E^0_j\right) \pi^U_t f_i
\label{eq:calcet}
\end{equation}
where $\pi^U_t$ is the fraction of maximum traffic of an user for a
period $t$.
We randomize each test point traffic with a 
a Gaussian distribution  $f_i$ with a mean of 1 and a scale of 0.2
truncated to the interval $[0,2]$.  
\subsection{Network Parameters}
\label{sec:net-def}
The small network  shown in Figure~\ref{fig:net}  is made up of 4
micro base stations~\cite{auer12} and~12 test points. These base stations need
94 and 39 watts of  power  in the  active and idle states. 
We can see that 
the coverage areas of the base stations need to overlap to some extent
to be able  to reassign some  of the test points to different
base stations when some base stations are in sleep mode.  

The variation of solar illumination and traffic demand over the day is
modeled by the parameters $\pi^S_t$ and $\pi^U_t$. They represent the
fraction of the peak value that is present in period $t$. They have
been taked from~\cite{marsan13}.

The illumination profile used to model the harvested solar
energy is that of the city of Palermo, Italy. The industrial grid cost is
set at 0.22\$/kWh which is representative of this country's real
electricity pricing~\cite{grave16}. The total
cost of a solar system calculated with~\eqref{eq:csj} is
\$2197 for the micro base station. This is an underestimation because
the cable and labor costs of the installation are not considered.
Note also that we can compute the total energy produced by the panels
based on the solar profile and the other panel parameters. In the
present case, this yields a value of 14 MWh over a 20-year
horizon. This yields an equivalent cost of \$0.16/Kwh, which is lower
than the grid cost. Based on these values, one might conclude that
installing solar panels everywhere should be the most economical
solution. As we will see later, it turns out that this is \emph{not}
the best solution due to the use of sleep mode.
\subsection{Solar Equipment Sizing}
\label{sec:pv_dim}
Our model does not optimize the number of solar panels and batteries
that are to be installed on the base stations. Instead, these are
chosen at the outset based on the following considerations.
\begin{table*}[]
  \centering
  \begin{tabular}{ccccccccc}
    Nb PVs  & TPs & \begin{tabular}[c]{@{}c@{}}Total \\
    	BSs
    \end{tabular} & \begin{tabular}[c]{@{}c@{}}Solar \\
    	BSs
    \end{tabular} & Cost (k\$) & Solar (k\$) & Grid
                                                     (k\$)
    & \begin{tabular}[c]{@{}c@{}}Solar used \\
        (MWh)
      \end{tabular}
            & \begin{tabular}[c]{@{}c@{}}Installed \\
                solar (MWh)
              \end{tabular} \\
    \hline
    3     & 12 & 4 & 0  & 12.55 & 0    & 12.5 & 0     & 0     \\
    4     & 12 & 4 & 4  & 12.21 & 7.68 & 4.53 & 36.46 & 37.41 \\
    5     & 12 & 4 & 4  & 10.43 & 8.13 & 2.3  & 46.58 & 46.76 \\
    6     & 12 & 4 & 4  & 9.995 & 8.79 & 1.21 & 51.55 & 56.11 \\
    7     & 12 & 4 & 3  & 10.12 & 7.08 & 3.04 & 43.23 & 49.1  \\
    8     & 12 & 4 & 3  & 11.44 & 8.42 & 3.03 & 43.28 & 56.11 \\
    9     & 12 & 4 & 3  & 11.75 & 8.75 & 3    & 43.40 & 63.12 \\
  \end{tabular}
  \caption{Solar Panels Sizing}
  \label{tab:pvs}
\end{table*}
First we try to estimate a good value for the number of solar panels
that should be installed on a base station. We vary the number of
panels and compute an optimal solution of problem PL in each case. These results
are computed for a time base 
of 24 slots which yields the more accurate results as discussed in
section~\ref{sec:quant-time}. 
We present in
Table~\ref{tab:pvs} both the solar and grid costs of the network over 
20 years.  In that period, the base stations will need 57.04 MWh of
electricity.

When the base stations are equiped with only 3 solar panels, the best
solution is not to  install any  because they produce too little
energy as compared with their cost. With 4 and 5  panels per base station, the
solution is to install 
solar equipment  everywhere and most of the harvested energy
is stored in the batteries without losses. This is because the
effective solar energy that is used to power the base stations, shown in column
\emph{Solar Used} of Table~\ref{tab:pvs}, is close to the
total solar energy that could be generated, as shown in the last
column \emph{Installed}. We see that the minimum cost is achieved with
6 solar panels. If the base
stations are equipped with more panels, the optimal solution is to
deploy solar equipment 
only in 3 of the 4 base stations. If we install more than this, the
total energy produced produced by the panels  will be larger than what
is actually needed and some of it will be lost.
The solar energy actually used becomes stable at a value
close to 43 MWh 
because the losses just keep accumulating with the increase of
available energy.  

The same procedure has been applied for the batteries and we concluded
that, for a set of batteries at different capacity and price, a
single 2568 VAh battery is enough to store the solar energy without
being too expensive. 

Therefore, in all the following, we  assume that base stations have 6
solar panels and a single  2568 VAh battery. The number of
inverters and charge controllers is set so that their nominal
power is greater or equal than the power of the solar panels and
batteries.

\begin{table*}[]
	\centering
	\begin{tabular}{lcccc}
		\multicolumn{1}{c}{Time Base} & TPs & BSs & Time (sec)
		& Cost (k\$) \\
		\hline
		0,\ 3,\ 6,\ 9,\ 12,\ 15,\ 18,\ 21 & 27 & 9 & 0.28 & 24.70  \\
		0,\ 1,\ 2,\ ...,\ 21,\ 22,\ 23 & 27 & 9 & 0.66 & 22.97  \\
		0,\ 9,\ 10,\ 13,\ 15,\ 18,\ 19,\ 20 & 27 & 9 & 0.1 & 22.97
	\end{tabular}
	\caption{Quantization of the Time Base}
	\label{tab:time-base}
\end{table*}
\subsection{Time Quantization}
\label{sec:quant-time}
We now define a good time base to use in our model. The
number of slots of the time base has three important effects. First,
having a larger number of slots means that the network can adapt more
accurately to the changes in demand or illumination, which will lead
to less costly solutions. On the other hand, switching the base
stations on and off can reduce the lifetime of the equipment so that
the smaller the number of transitions the better. Also, from the point
of view of calculating a solution, a larger time base increases the
number of variables and constraints and will increase the solution time.

In order to estimate the effect of the time base, we first consider
two cases with uniform interval sizes, one with one-hour and the other with
three-hour intervals. 
We then solve problem PL for a medium sized network of 27 test points and 9 base stations with these two time 
bases. The results are presented in the first two rows of
Table~\ref{tab:time-base}. From this, we can see that having smaller
time slots does improve the quality of the solution as expected but
takes longer to solve. The first time base
has 8 time slots and takes
0.28 second to be solved to optimality with a total cost 
of 24.7~k\$. With 24 time slots, the optimization takes
0.66 second but has a lower cost of 22.97~k\$.

\begin{figure}
  \centering
  \includegraphics[scale=0.5]{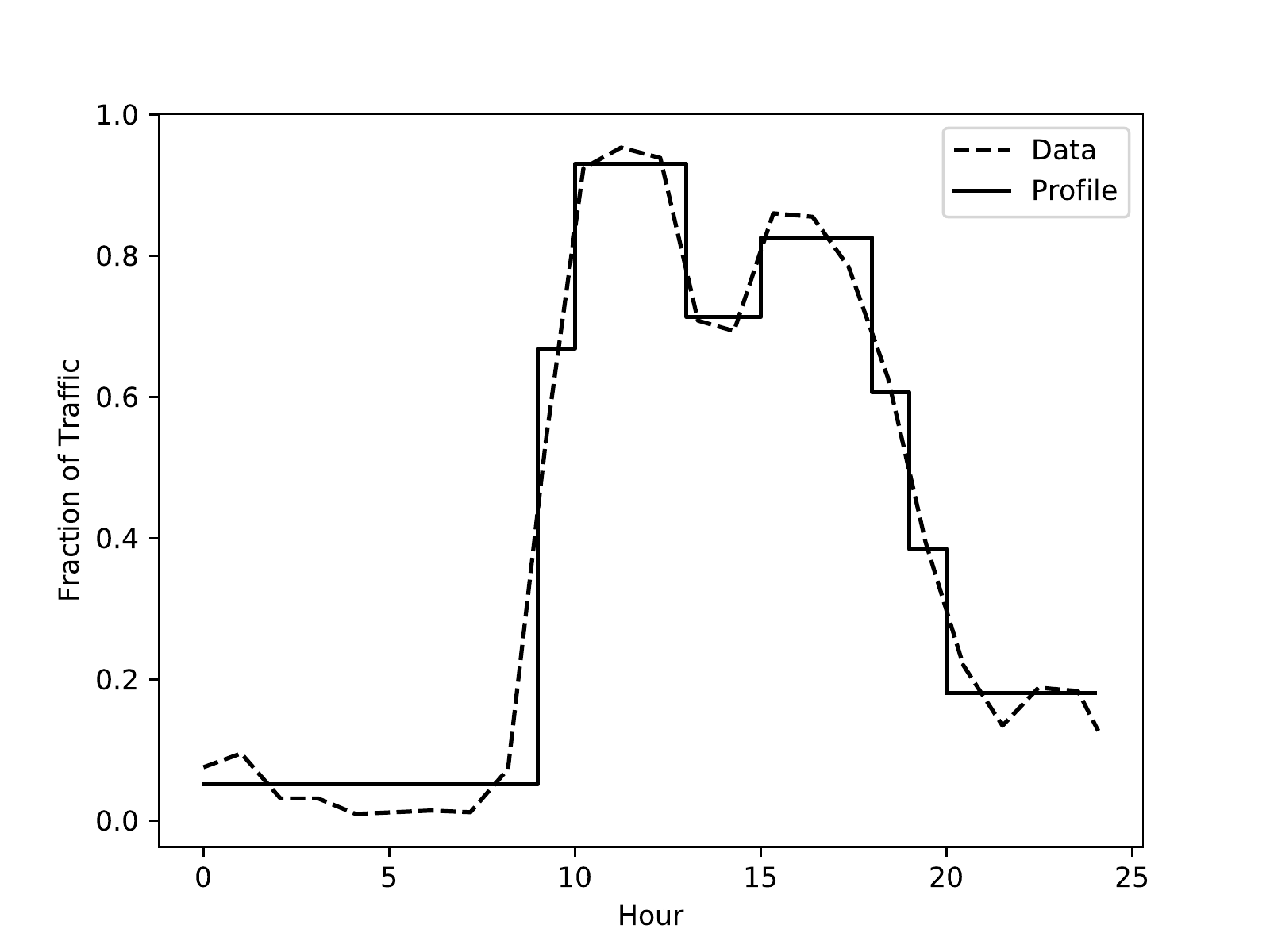}
  \caption{Traffic Profile}
  \label{fig:traffic}
\end{figure}
\begin{figure}
  \centering
  \includegraphics[scale=0.5]{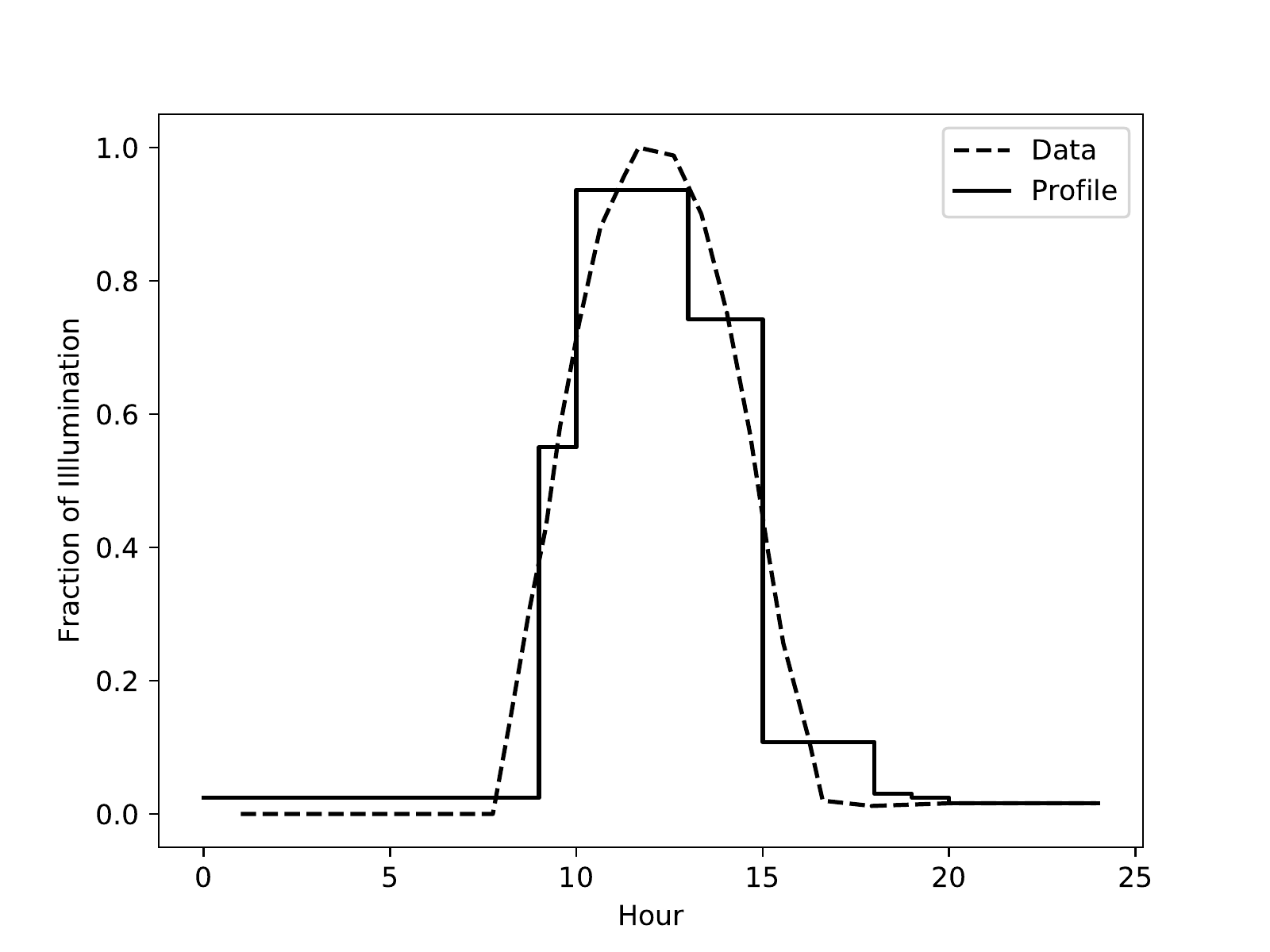}
  \caption{Illumination Profile}
  \label{fig:illumination}
\end{figure}

We can strike a compromise between the computation time and the
quality of the solution if we choose a time base better suited to the
actual  traffic and solar radiation
profiles. We can then choose the number of time slots desired and do a
best fit of the slot sizes with respect to the actual profiles.
We can see in figures~\ref{fig:traffic} and~\ref{fig:illumination}
the traffic and illumination profiles along with the fitted time
base shown in~\eqref{eq:time_base}.
\begin{equation}
T = \{ 0,9,10,13,15,18,19,20 \} . 
\label{eq:time_base}
\end{equation}
This time base is relatively small and we can see from the third line
of Table~\ref{tab:time-base} that the cpu time of
0.1 seconds is much smaller than that of the 24-slot base with a
total cost of 22.97~k\$, which is the same cost as with 24 time slots. 

Based on these results, we will be using  8 time slots for the results
of~section~\ref{sec:performance}. For the results
of~\ref{sec:dyn-net}, we use 24 time slots to
have a clearer view of the changes in the solution for the different
optimizations.  
\section{Network Results}
\label{sec:network-results}
We now present more detailed results for six networks of increasing
size. In each case, we consider a number of scenarios to look at the
interaction between the solar installation and the use of sleep
mode. Each scenario requires the solution of a 
variation of problem PL where some variables are fixed to some value
and the optimization is on the set of remaining variables.
\subsection{Scenarios}
\label{sec:proposed_sol}
First we consider the {\em base} case where neither solar power nor sleep
mode is used.
This scenario corresponds to most present networks and can be
used as a  comparison point. The next
two scenarios correspond to the use of either sleep
mode or solar power but not both. We call these \emph{single
  technology} scenarios.
Then
we consider two scenarios for introducing both technologies in a
network, one after the other. We call these \emph{sequential} scenarios.
In both cases, the end result is a network using both solar
power and sleep mode. The difference is in the way we evolve the
current network to its final configuration.
The final scenario is when we plan the network with both technologies
available from the start. We call this the \emph{joint} scenario.
\subsubsection{Problem P1: Base Network}
\label{sec:nonewtech}

In this scenario, there is no solar equipment or sleep mode available and the
base stations are always 
on. We design the network by solving a restricted version of problem
PL where the 
variable $z_j$ and $x^o_{j,t}$ are fixed to 0 so that 
the only optimization is on the assignment variables $h_{i,j,t}$. 
\subsubsection{Problems P2 and P3: Single Technology}
\label{sec:singletech}
Problem P2 is the case 
when solar is not available so that we fix the $z_j = 0$ and optimize
over the other variables $x^o_{j,t}$ and $h_{i,j,t}$.
In case P3, when 
solar is available but not sleep mode, we fix the  $x^o_{j,t}
= 0$ and solve again for the remaining variables.
\subsubsection{Problem P4: Sequential  Scenario: Sleep Mode First}
\label{sec:sleepfirst}
In this first sequential scenario,
we design  the network in two steps by solving a different special case of
problem PL each time. 
At first,  we don't
use solar energy so that we fix the $z_j = 0$ and 
optimize the network over the scheduling variables  $x^o_{j,t}$ and the test
point assignment $h_{i,j,t}$.
Next, to model the introduction of solar equipment in the network,  we
solve a new special case of PL 
based on the results of the first step
where sleep mode has already been planned. For this, the variables
$x^o_{j,t}$ and 
$h_{i,j,t}$ are fixed at their current values and
the  optimal values of the  $z_j$ variables are recomputed by solving
this restricted version of PL. 
\subsubsection{Problem P5: Sequential Scenario: Solar First}
\label{sec:solarfirst}
In this scenario, the sequence
is inverted: solar power is introduced first so that  we fix the
$x^o_{j,t} = 0$ and optimize the placement of 
solar equipment via the $z_j$ variables. These variables and the test
point assignments $h_{i,j,t}$ are then
fixed at their current value and the sleep mode schedule $x^o$ is optimized.
\subsubsection{Problem PL: Joint Optimization}
\label{sec:jointopt}
Finally, we get to the scenario when both technologies are optimized at the same
time by solving the full problem PL described in
Section~\ref{sec:line-optim-model}. This is obviously the best option
and we can compare the benefit of optimizing  the two options \emph{at
  the same time} to the previous cases. 
\subsection{Solution Algorithm}
\label{sec:algo}
In this paper, 
problem PL and its variants are solved using
Gurobi with the default options and  the Ampl pre-processor. Once
reduced by Ampl, the largest network we have solved
has 42408 rows, 32959 columns and  125082 binary variables. 

For large cases,
we were unable to solve the joint PL problem from a cold start.
We found that Gurobi would generate a large search tree trying to find a
feasible solution and would run out of memory even on a 
machine with a large memory. Incidentally, the same thing happened
with the Cplex solver.
Nevertheless, it is possible to solve these cases to a reasonable
accuracy if one starts the optimization with a  known feasible
solution. This could be the solution of any one of the  P1 to P5
problems and in all the following, the solution of PL is always
computed with one of these solutions as the starting point. 
\begin{table}[]
  \centering
  \begin{tabular}{lllcc}
    Problem & Solar & Sleep & Cpu & Gap \\
    Name & Available & Available & & \%\\
    \hline
    \multicolumn{5}{|c|}{123 TP, 41 BS}\\
    \hline
P1 & no & \ no & 0.0183    & 0          \\
P2 & no & \ yes  & 41.2    & 0          \\
P3 & yes  & \ no & 0.0665    & 0          \\
P4 & yes  & \ yes* & 0.0564     & 0          \\
P5 & yes* & \ yes  & 1.28    & 0          \\
PL & yes  & \ yes  & 232    & 2       \\
            &     &             &                 \\
    \hline
    \multicolumn{5}{|c|}{216 TP, 72 BS}\\
    \hline
P1 & no & \ no & 0.0773    & 0          \\
P2 & no & \ yes  & 307    & 0.6       \\
P3 & yes  & \ no & 0.124    & 0      \\
P4 & yes  & \ yes* & 0.089      & 0    \\
P5 & yes* & \ yes  & 26.8     & 0    \\
PL & yes  & \ yes  & 187    & 3        \\
             &     &              &                 \\
    \hline
    \multicolumn{5}{|c|}{486 TP, 162 BS}\\
    \hline
P1 & no & \ no & 7.39      & 0          \\
P2 & no & \ yes  & 27454   & 0.3       \\
P3 & yes  & \ no & 0.411    & 0          \\
P4 & yes  & \ yes* & 0.271    & 0    \\
P5 & yes* & \ yes  & 310     & 0     \\
PL & yes  & \ yes  & 14067   & 3     \\
&     &              &        \\
\hline
\multicolumn{5}{|c|}{864 TP, 288 BS}\\
\hline
P1 & no & \ no & 132      & 0          \\
P2 & no & \ yes  & 109058   & 0.5       \\
P3 & yes  & \ no & 0.815    & 0          \\
P4 & yes  & \ yes* & 0.442    & 0    \\
P5 & yes* & \ yes  & 5560     & 0.01     \\
PL & yes  & \ yes  & 54352   & 4           
  \end{tabular}
  \caption{CPU time (sec). * indicates option available first}
  \label{tab:big-net-1}
\end{table}
The cpu time is the total time used by all of the processors. 
Some values for the three larger networks are shown in
Table~\ref{tab:big-net-1} along with the optimality gap at the final
solution.
The first column shows which of the restricted problems  is being solved.
The \emph{Solar} and \emph{Sleep} headings  indicate whether whether a
technology, solar or sleep mode, is available (yes) or not (no). 
For the sequential scenarios of Sections~\ref{sec:sleepfirst}
and~\ref{sec:solarfirst},  a \emph{starred} entry means that 
this was the first technology introduced.
Finally, the remaining columns show the total cpu time needed and the
relative optimality gap.
It is quite clear from these results that using Gurobi is a realistic
option for off-line dimensioning of networks of up to  300 base
stations. Larger networks will require heuristic
techniques specifically taylored to the problem.
\begin{table}
  \centering
  \begin{tabular}{llcc}
    & & Sleep First & Solar First \\
    & & P4 & P5 \\
    \hline
    \multicolumn{4}{c}{12 TP, 4 BS} \\
    \hline
    P1 & Base & \multicolumn{2}{c}{14.5}\\
    & Step 1 & 12.4 & 11.2 \\
    & Step 2 & 10.1 & 10.0 \\
    PL & Joint & \multicolumn{2}{c}{9.7}\\
    \hline
    \multicolumn{4}{c}{54 TP, 18 BS} \\
    \hline
    P1 & Base & \multicolumn{2}{c}{65.2}\\
    & Step 1 & 54.4 & 50.4 \\
    & Step 2 & 45.0 & 43.2 \\
    PL & Joint & \multicolumn{2}{c}{42.8}\\
    \hline
    \multicolumn{4}{c}{123 TP, 41 BS} \\
    \hline
    P1 & Base & \multicolumn{2}{c}{149}\\
    & Step 1 & 118 & 115 \\
    & Step 2 & 99.8 & 98.5 \\
    PL & Joint & \multicolumn{2}{c}{95.3}\\
    \hline
    \multicolumn{4}{c}{216 TP, 72 BS} \\
    \hline
    P1 & Base & \multicolumn{2}{c}{261}\\
    & Step 1 & 205 & 202 \\
    & Step 2 & 175 & 166  \\
    PL & Joint & \multicolumn{2}{c}{164}\\
    \hline
    \multicolumn{4}{c}{486 TP, 162 BS} \\
    \hline
    P1& Base & \multicolumn{2}{c}{587}\\
    & Step 1 & 454 & 454 \\
    & Step 2 & 390 & 367  \\
    PL & Joint & \multicolumn{2}{c}{362}\\
    \hline
    \multicolumn{4}{c}{864 TP, 288 BS} \\
    \hline
    P1& Base & \multicolumn{2}{c}{1044}\\
    & Step 1 & 791 & 807 \\
    & Step 2 & 683 & 646  \\
    PL & Joint & \multicolumn{2}{c}{646}
  \end{tabular}
  \caption{Optimal network cost (k\$) for sequential algorithms:
    Effect of order}
  \label{tab:compare}
\end{table}
\subsection{Sequential Scenarios}
\label{sec:performance}
We now examine in Table~\ref{tab:compare} the effect of sequential scenarios where we optimize
one technology after the other.
As before,
the results for each network are grouped in blocks of five lines. The
first boxed line shows the number of test points and base stations and
the next four one, the total cost 
for each scenario. The first of these lines is the cost of the base
configuration obtained by solving P1  without sleep mode or solar
power. The next two lines 
represent the two scenarios for sequential optimization P4 and P5. The column
marked \emph{Sleep First} is for the case where the sleep mode is
optimized first. In that case, \emph{Step~1} is to optimze the sleep
mode and \emph{Step~2}, the solar installation.
For  column, labelled \emph{Solar First},
solar installation is optimized first, which is \emph{Step~1}, and
then sleep mode, in \emph{Step~2}. Finally, the last line of
the block shows the total cost when both options are optimized
together by solving PL.

An interesting point is that order \emph{does} matter when doing
sequential optimization. For all cases tested, using
solar equipment before the sleep schedule yields better results than
doing it in the opposite order. 

Another conclusion is that the results of problem P5, where solar
equipment is optimized first, are
very close to those of the joint optimization PL, which is the truly
optimal solution. 
A simple and accurate heuristic could then be to 
optimize first the solar installation and then add the dynamic
operation of base stations without having to do the joint scenario PL.

Finally, starting from row~1 to~4, we can see from the results that
the savings  offered by the two technologies are clearly additive
irrespective of the order in which they are planned. 
\begin{table*}[]
	\centering
	\begin{tabular}{lllccccccc}
         \begin{tabular}[l]{@{}c@{}}Pb\\
                                      No\end{tabular}  & 
         \begin{tabular}[l]{@{}c@{}}Solar\\
                                      Avail\end{tabular} &
         \begin{tabular}[l]{@{}c@{}}Sleep\\
                                      Avail\end{tabular} &
                                    \begin{tabular}[c]{@{}c@{}}Cost\\
                                      (k\$)\end{tabular}
                & \begin{tabular}[c]{@{}c@{}}Total\\Solar cost\\
                    (k\$)\end{tabular}
                & \begin{tabular}[c]{@{}c@{}}Grid\\
                    (k\$)\end{tabular}
                & \begin{tabular}[c]{@{}c@{}}Unit\\Solar cost\\
                    (\$/kWh)\end{tabular}
                & \begin{tabular}[c]{@{}c@{}}Solar \\used 
                    \end{tabular}
                & \begin{tabular}[c]{@{}c@{}}Inst \\
                    solar \end{tabular}
                & \begin{tabular}[c]{@{}c@{}}Ant\\
                    energy \end{tabular} \\
          \hline
P1  &		\ \ no   & \ \ \ \ no     & 14.49     & 0     &
                                                                  14.49  & NA     & 0      & 0      & 65.9    \\
   P2       &
		\ \ no   & \ \ \ \ yes      & 12.55     & 0     &
                                                                  12.55  & NA     & 0      & 0      & 57    \\
     P3     &
		\ \ yes    & \ \ \ \ no     & 11.2      & 8.79  & 2.42
                & 0.16   & 54.9   & 56.1   & 65.9    \\
       P4   &
		\ \ yes    & \ \ \ \ yes*     & 10.02     & 6.59  & 3.43
                & 0.159  & 41.5   & 42.1   & 57    \\
          P5 &
		\ \ yes*    & \ \ \ \ yes     & 9.995     & 8.79  & 1.21
                & 0.169   & 51.6   & 56.1   & 57    \\
          PL &
		\ \ yes    & \ \ \ \ yes      & 9.995     & 8.79  & 1.21   & 0.169   & 51.6   & 56.1   & 57  
	\end{tabular}
	\caption{Solar use, small network, energy in MWh}
	\label{tab:small-net-24t}
      \end{table*}

\subsection{Solar Energy}
\label{sec:soluse}
Here, we discuss the results regarding the use of solar energy in the different networks. The installed and used solar energy is examined for each of the optimization problems. In section~\ref{sec:detsol}, it is done for the small network and section~\ref{sec:optimpactsol} presents a summary of the results for the bigger networks. 
\subsubsection{Detailed Solar Use for the Small Network}
\label{sec:detsol}
To demonstrate how the network makes use of solar energy, we present in
Table~\ref{tab:small-net-24t} more results for the small network
of Figure~\ref{fig:net}
with 12 test points and 4 base stations, a 24-slot time base and where all the solutions
are optimal. For this small network, the results from P5 and PL are
identical. It is not common to have an optimal solution when the solar
installation is solved first but it might happen in small networks.  

In that table, the last column labeled  \emph{Ant energy}
is the total energy needed by all the antennas over the study
period. It depends on the total energy required by the users and
how many base stations are in sleep mode.
The  \emph{Inst solar} column is the total amount of energy
that the installed solar equipment can produce
while the column  \emph{Solar used} is the amount of solar energy that
was effectively used by the network. 
The difference between the \emph{Inst
  solar}  and \emph{Solar used} columns is
the amount of lost energy due to battery overflow~\eqref{eq:cycle}.

In that table, problems P1 and P3 correspond to the two cases where there
is no sleep mode. As expected, they have a larger energy requirement than
the other four cases where sleep mode is used. Out of
these four cases, P3 is the one that uses the largest amount of
solar energy. Here, the antennas need the most energy and sleep mode 
is not available so that 
the solution is to install as much solar equipment
as possible, leading to a large solar usage (col Solar used)  with a correspondingly
large capital cost (col Total solar cost).

Conversely, the smallest solar usage is that of P4,  when sleep
mode is planned 
first, without solar, and then the solar equipment is installed. Using
sleep mode reduces considerably the amount of energy needed by the
antennas (col Ant energy) but the amount of solar energy effectively used
(col Solar used) also
decreases significantly, with a corresponding decrease in the capital
cost (col Total solar cost). The downside is that because there is less solar equipment
available, the grid cost increases (col Grid) so that the total cost
(col Cost) is still relatively large.

The important point to note is that \emph{neither} of these two
solution is optimal  and the solution that uses the most solar energy
is definitely 
\emph{not} the best one. The solutions of problems P5 and PL strike a
balance between the reduction of grid energy on the one hand and the
ensuing capital cost and yield a value for the amount of
solar energy used that is between the ones of P3 and P4. This shows
that any solar optimization 
technique must take into account the capital cost of the solar
equipment before deciding to deploy this technology and that one
cannot assume that it comes for free.

In column \emph{Unit Solar Cost}, the solar price
in dollars per kilowatt hour is simply the total cost of the solar
equipment installed divided by the amount of solar energy that was
used. This turns out to be 0.169\$/kWh for the best solutions of
P5 and PL, which is 
smaller  than  the 0.22\$/kWh grid cost so that solar is an
economically viable option.

Similar results hold for the other networks. A summary of the results
for the larger networks is shown in section~\ref{sec:optimpactsol}. 

\begin{figure}
	\centering
	\includegraphics[scale=0.5]{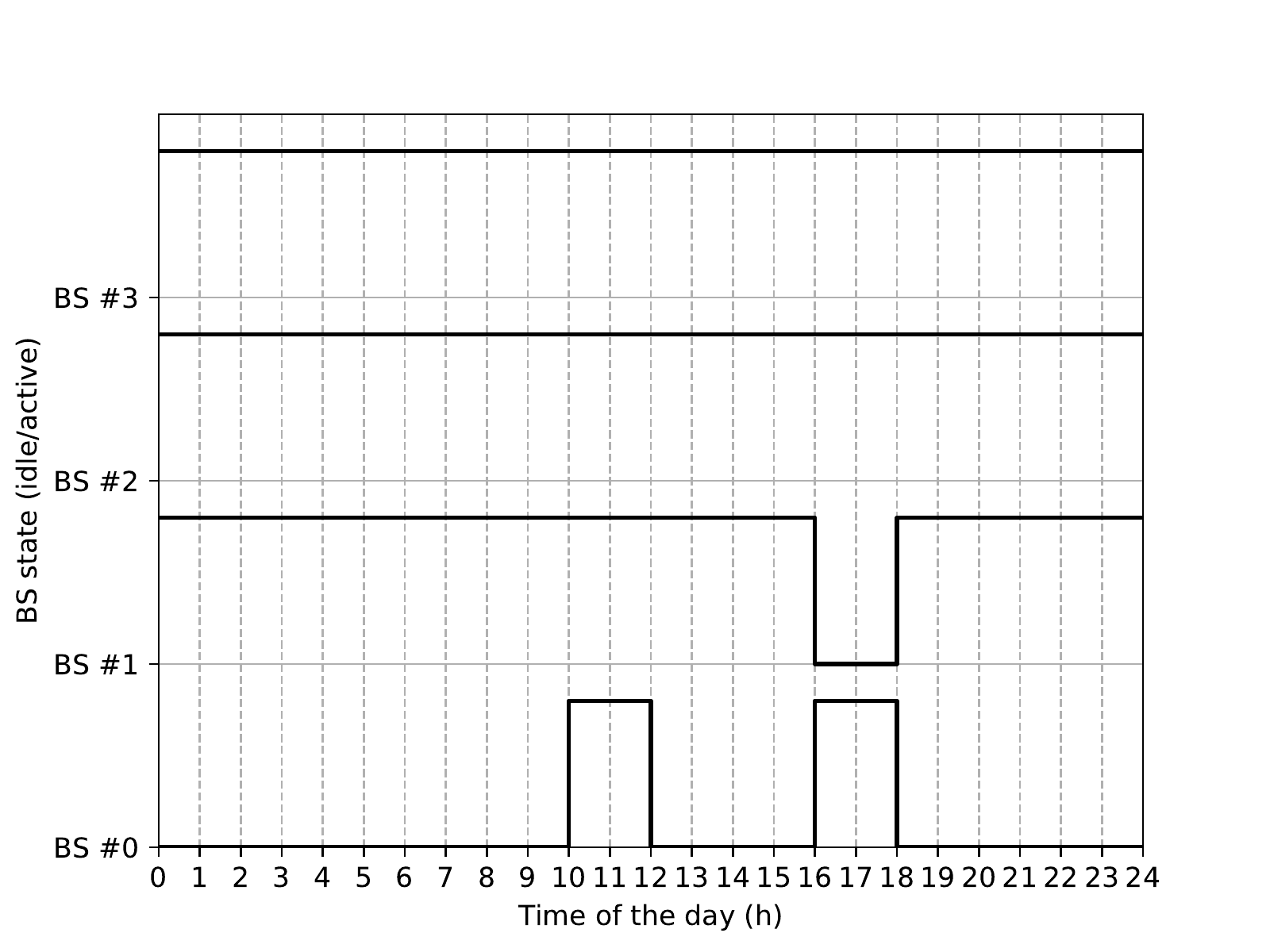}
	\caption{base stations activation, sleep optimization first}
	\label{fig:bs-sol-fix-dyn}
\end{figure}

\begin{table}[]
	\centering
	\begin{tabular}{ccccc}
          Pb & Solar & Total Cost & Solar Energy \\
          No & BSs & (k\$) & Losses (\%) & \\
		\hline
		\multicolumn{5}{|c|}{54 TP, 18 BS}\\
		\hline
		PL &		17  &  42.8 & 4.2 \\
		PFS &		18 &  43.1 & 4.3 \\
		&     &             &                 \\
		\hline
		\multicolumn{5}{|c|}{123 TP, 41 BS}\\
		\hline
		PL &		34  &  95.9 & 3.7 \\
		PFS &		41 &  98.5 & 4.7 \\
		&     &             &                 \\
		\hline
		\multicolumn{5}{|c|}{216 TP, 72 BS}\\
		\hline
		PL &		65  &  164 & 4.9 \\
		PFS &		72 &  166 & 5.9 \\
		&     &              &                 \\
		\hline
		\multicolumn{5}{|c|}{486 TP, 162 BS}\\
		\hline
		PL &		158  &  362 & 5.2 \\
		PFS &		162 &  367 & 6.4  \\
		&     &              &                 \\
		\hline
		\multicolumn{5}{|c|}{864 TP, 288 BS}\\
		\hline
		PL &		288  &  646 & 6.8 \\
		PFS &		288 &  646 & 6.8
	\end{tabular}
	\caption{Installed Solar base stations}
	\label{tab:instsolbs}
\end{table}

\begin{figure}
	\centering
	\includegraphics[scale=0.5]{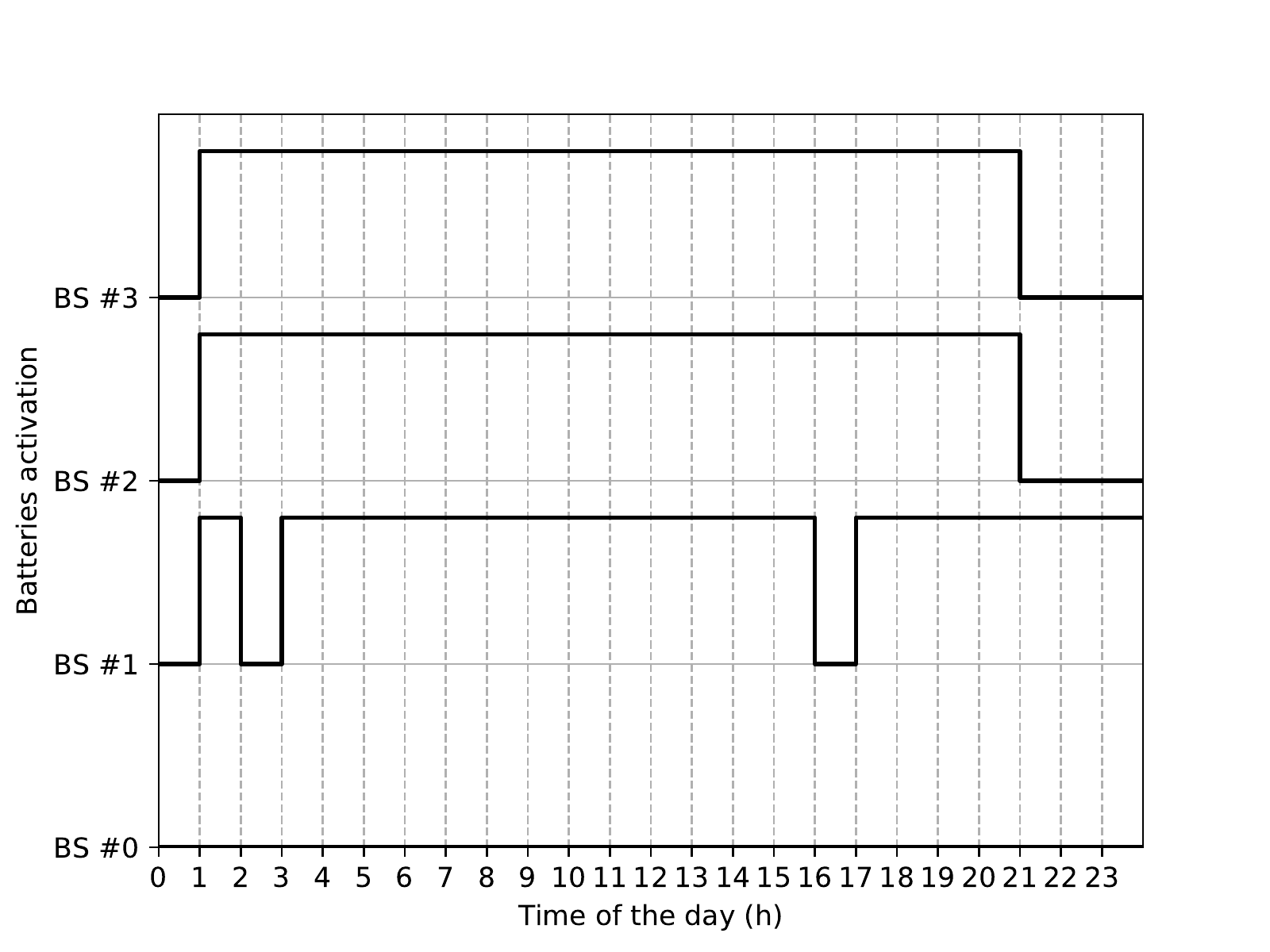}
	\caption{Batteries activation, sleep optimization first}
	\label{fig:batt-sol-fix-dyn}
      \end{figure}

\subsubsection{Optimization and Solar Installation}
\label{sec:optimpactsol}
We now present other results regarding the installation of solar
equipment on the base stations. The goal here is to show that
installing solar everywhere need not be the best solution  even
when  solar
energy is cheaper than grid energy. For this purpose, we use a model
where solar is installed everywhere. This is  
called problem PFS, a variant of PL where the variable $z_j$ is set to 1.

Table~\ref{tab:instsolbs} shows the number of base stations that have
solar installed in the column \emph{Solar base stations} where proble PFS has
been solved to optimality. We see that solar is
installed on most, but not all base stations even though
solar energy is 
cheaper than the grid. This is still true for bigger networks where the solar
energy rates are between 0.16\$/kWh and 0.17\$/kWh depending on the
network.  

Further information is provided in column \emph{Total Cost} where the
objective function value is shown. For the first three networks, the total cost
is smaller when some base stations do not use solar energy. For the
two larger networks, the results are not conclusive  because the
solutions are not optimal, with  gaps of 3\% and 4\%. 

At last, column \emph{Solar Energy Losses} presents the energy that is
lost due to the limited capacity of the batteries. It is a ratio of the
energy that is effectively used to power the base stations over the total
installed solar energy. This result shows that having too much solar
equipment in the network can lead to more losses.
\subsection{Network Dynamics}
\label{sec:dyn-net}
We now examine
in detail how the presence or absence of either solar or sleep mode
can affect the operation of the network both from the point of view of
the activation of the base stations and of the use of solar
energy. For this, we consider the two scenarios of
Sections~\ref{sec:sleepfirst} and~\ref{sec:solarfirst} where the small
network is optimized sequentially.
\subsubsection{Sleep Optimization First}
\label{sec:bs-act}
First we examine the case where we optimize sleep mode  first
and then solar.
We can see in 
Figure~\ref{fig:bs-sol-fix-dyn}  the activation of the base stations during
the day.

Base stations No~2 and~3 are always on
because they need to serve some test points that are only covered by
them. Base station~0 is almost always in the sleep mode except
during the two traffic peaks from 10:00 to 12:00 and 16:00 to 18:00,
as shown in Figure~\ref{fig:traffic}.
Base station~1 is almost always on except during the peak period at
16:00 where the traffic is taken up by base station~0.

This behavior is to be compared with the use of solar energy shown in
Figure~\ref{fig:batt-sol-fix-dyn}. We see that base station~0 never
uses solar while~1,~2 and~3 use solar almost all of the time. This is
a direct consequence of the optimization procedure. Because we optimze
the sleep schedule first without solar power, the network has fewer
opportunities to turn off the base stations and tries to compensate by
using as much solar power as possible. This is consistent with the
results of Table~\ref{tab:small-net-24t}.

\subsubsection{Solar Optimization First}
\label{sec:batt-act}
Next, we consider the case where the network is optimized first for
solar equipment and then for the sleep schedule. In the present case,
this solution is also the optimal solution where both technologies are
optimized together.

\begin{figure}
	\centering
	\includegraphics[scale=0.5]{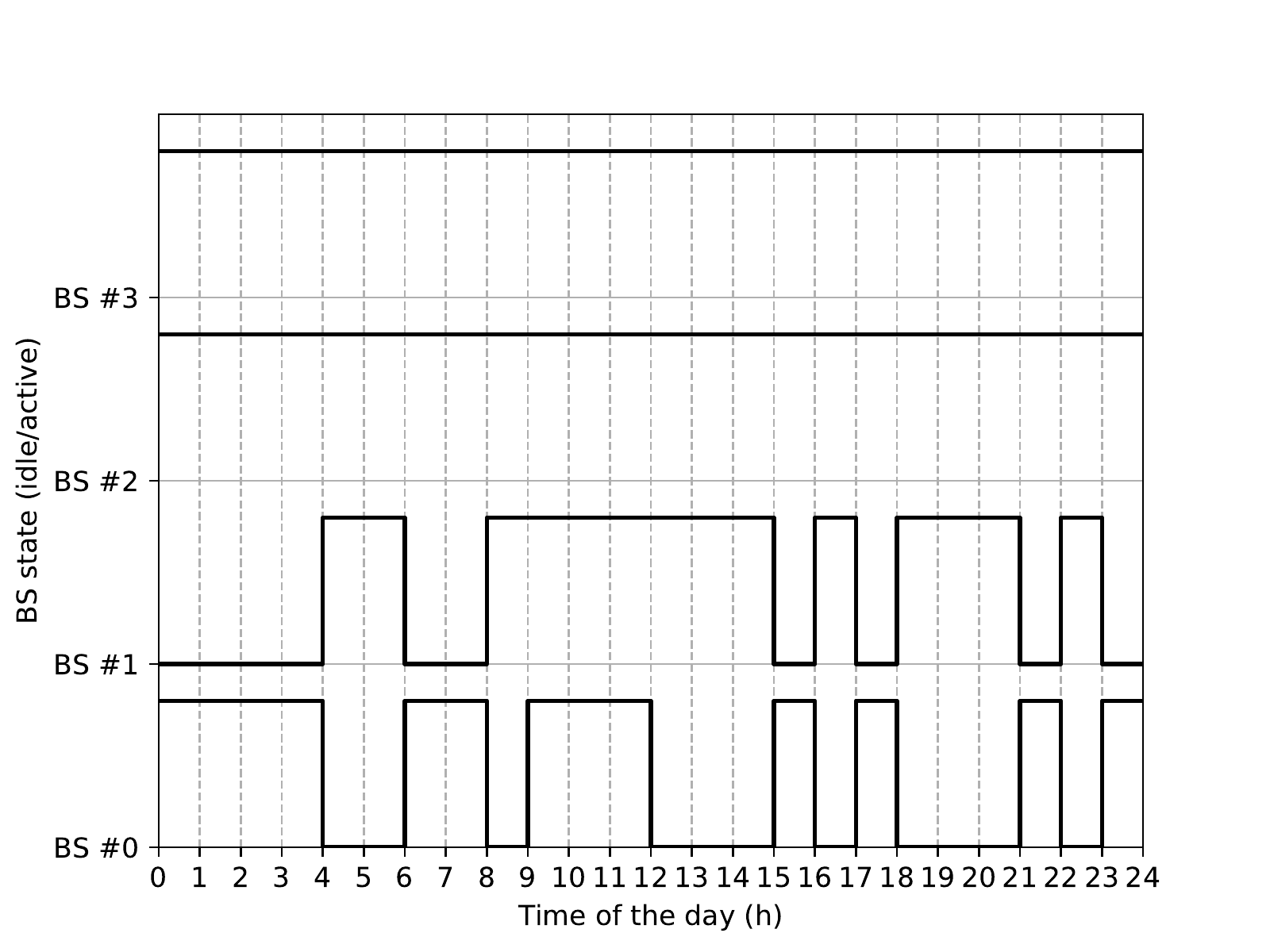}
		\caption{base stations activation, solar optimization first}
	\label{fig:bs-fix-sol-dyn}
      \end{figure}
\begin{figure}
	\centering
	\includegraphics[scale=0.5]{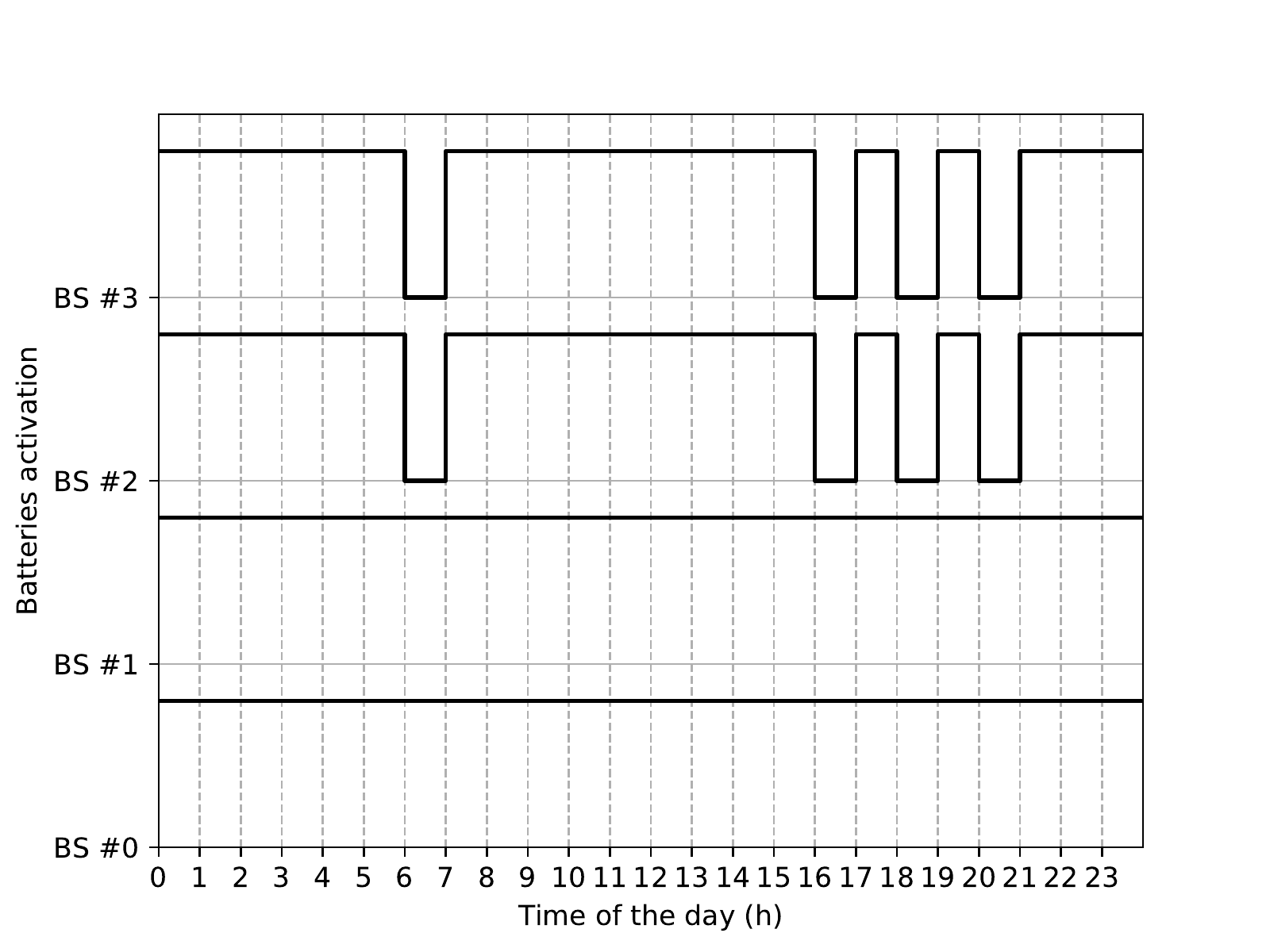}
	\caption{Batteries activation, solar optimization first}
	\label{fig:batt-fix-sol-dyn}
      \end{figure}
      
We can see in Figure~\ref{fig:bs-fix-sol-dyn} the base station
activation schedule. This is strikingly different from
Figure~\ref{fig:bs-sol-fix-dyn}. Base stations~2 and~3 are still on
all the time but~0 and~1 use sleep mode much more often. 
Use of solar energy is also different. Base stations~2 and~3 use grid
energy more often while
base stations~0 and~1 can now use solar power all the
time for added savings.

\begin{figure*}
	\centering
		\begin{subfigure}[b]{0.45\textwidth}
			\includegraphics[width=\textwidth]{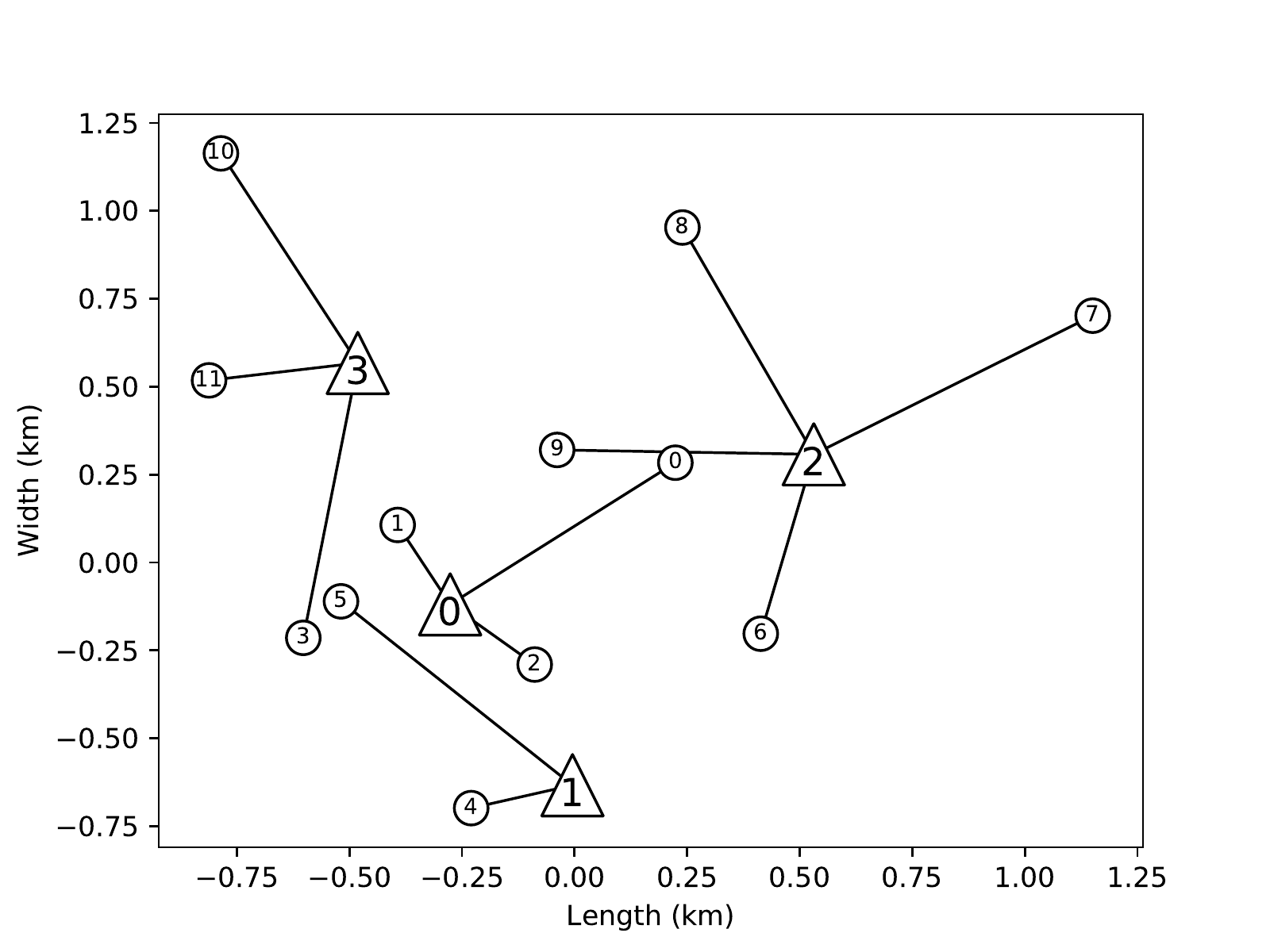}
			\caption{10:00}
			\label{fig:time-10}
		\end{subfigure}
		\begin{subfigure}[b]{0.45\textwidth}
			\includegraphics[width=\textwidth]{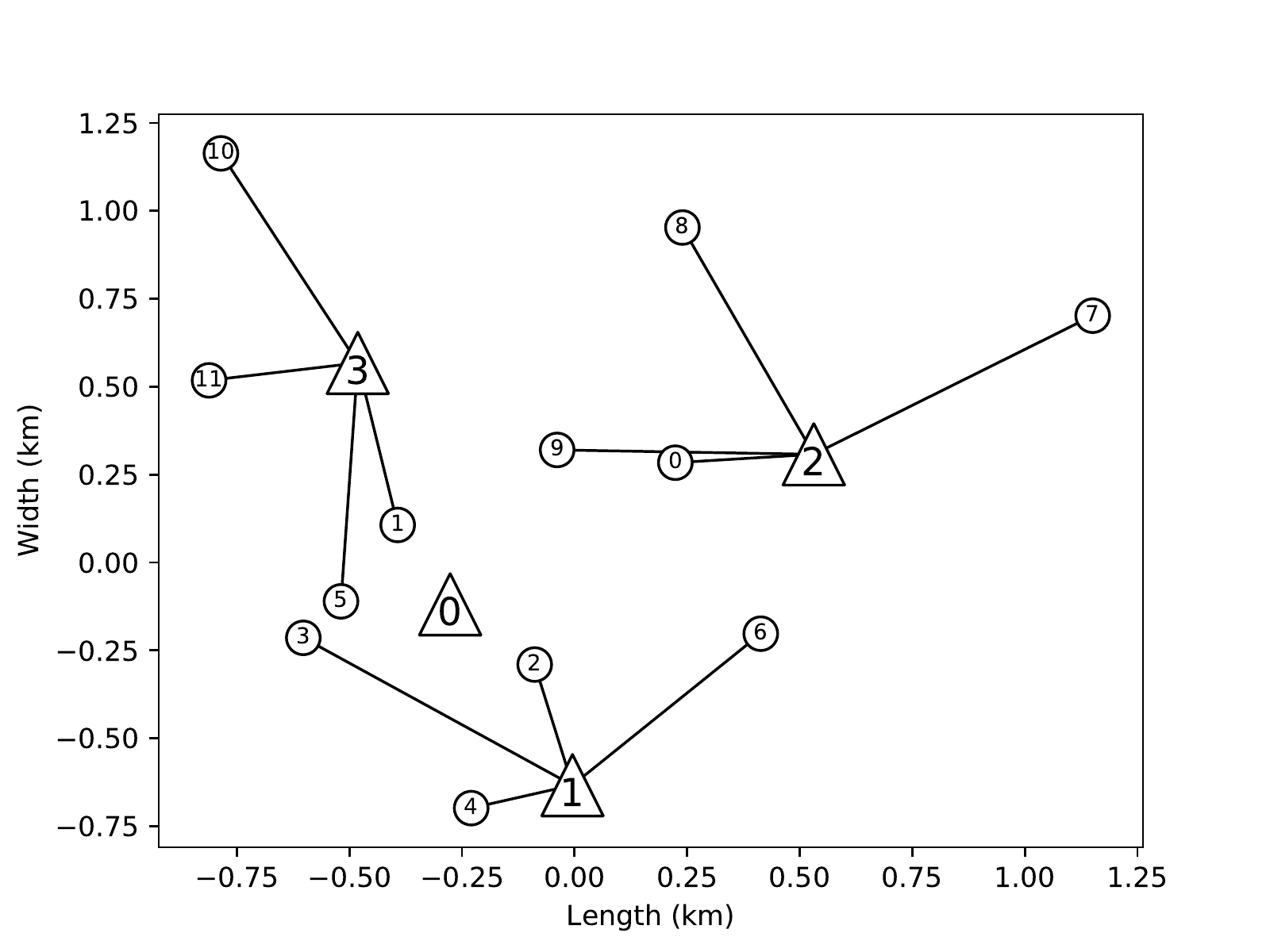}
			\caption{16:00}
			\label{fig:time-16}
		\end{subfigure}
		\begin{subfigure}[b]{0.45\textwidth}
			\includegraphics[width=\textwidth]{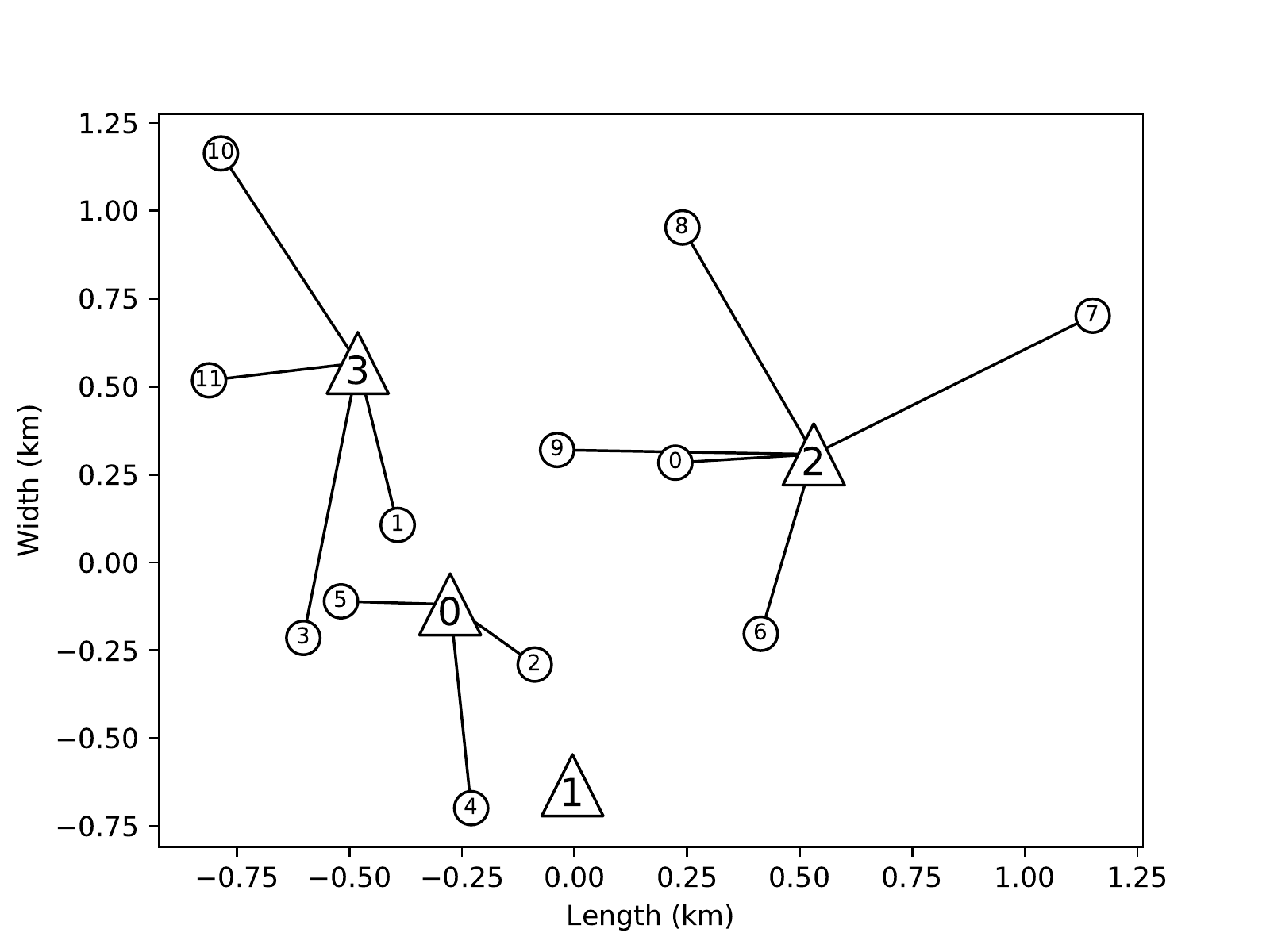}
			\caption{17:00}
			\label{fig:time-17}
		\end{subfigure}
		\begin{subfigure}[b]{0.45\textwidth}
			\includegraphics[width=\textwidth]{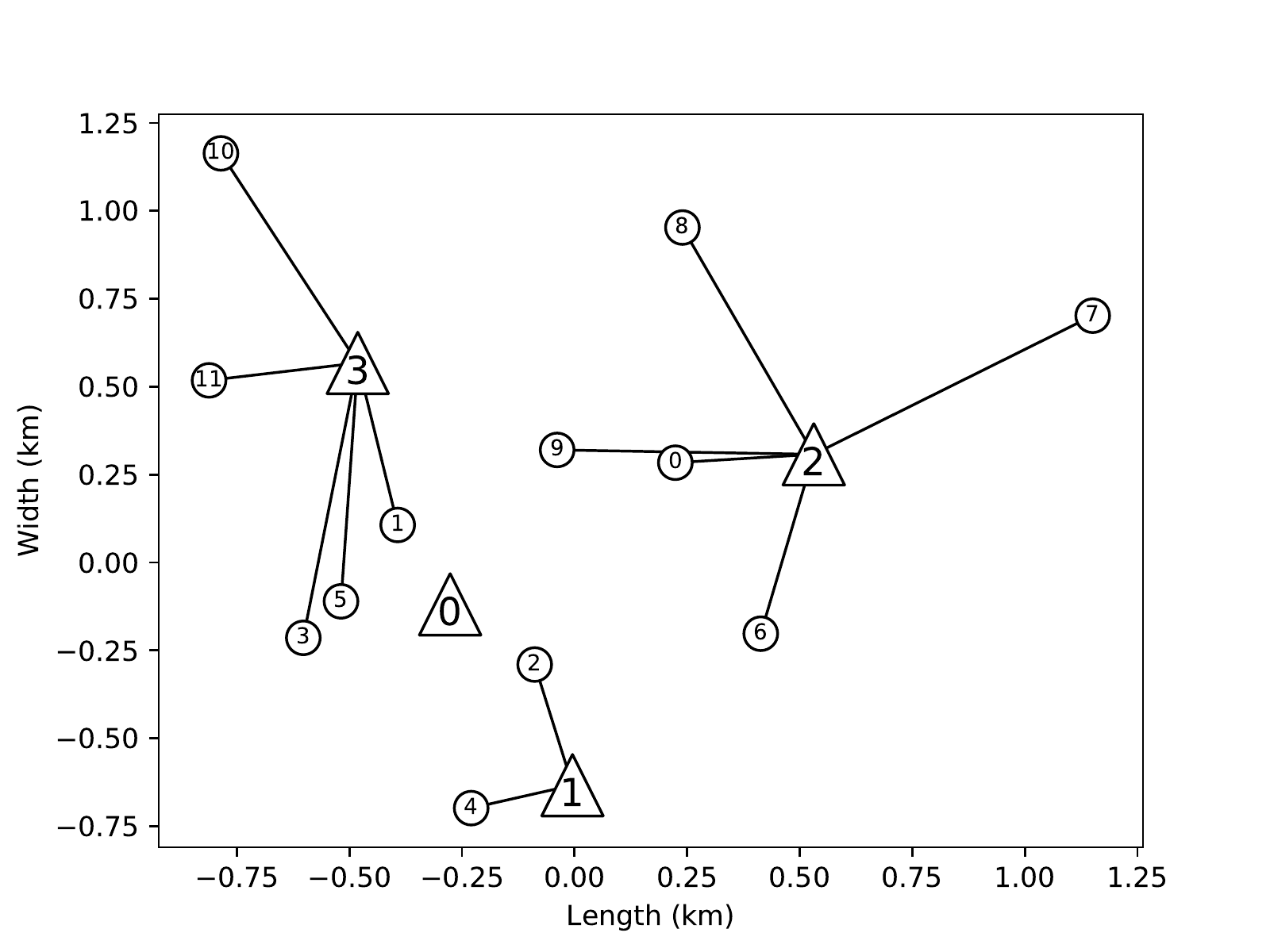}
			\caption{18:00}
			\label{fig:time-18}
		\end{subfigure}
		\caption{test points Assigments, solar optimization first}
	\label{fig:tps-assign}
\end{figure*}

\begin{figure}
	\centering
	\includegraphics[scale=0.5]{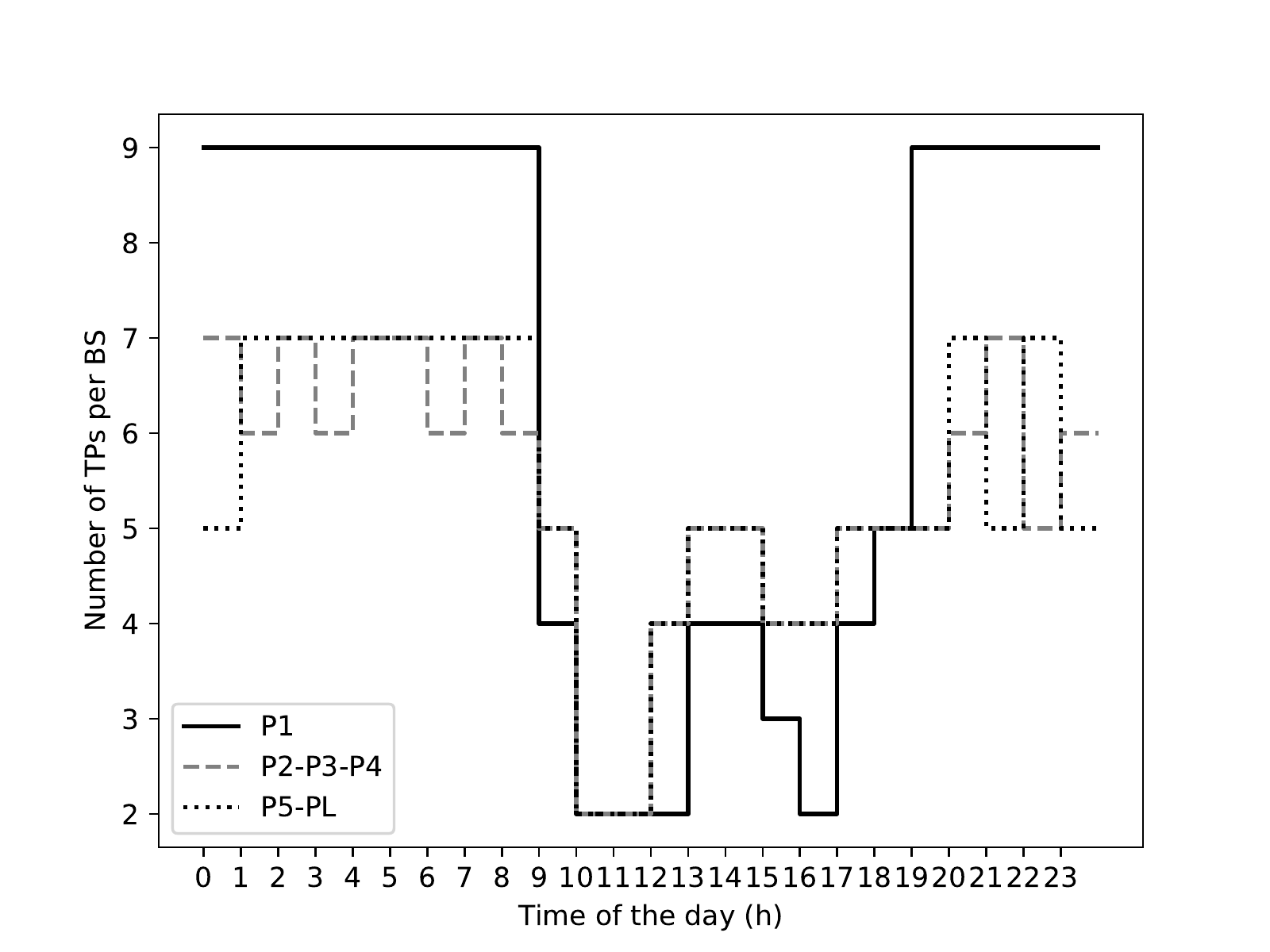}
	\caption{Dispersion Profiles}
	\label{fig:disp}
\end{figure}
      
\begin{figure}
    \centering
    \includegraphics[scale=0.5]{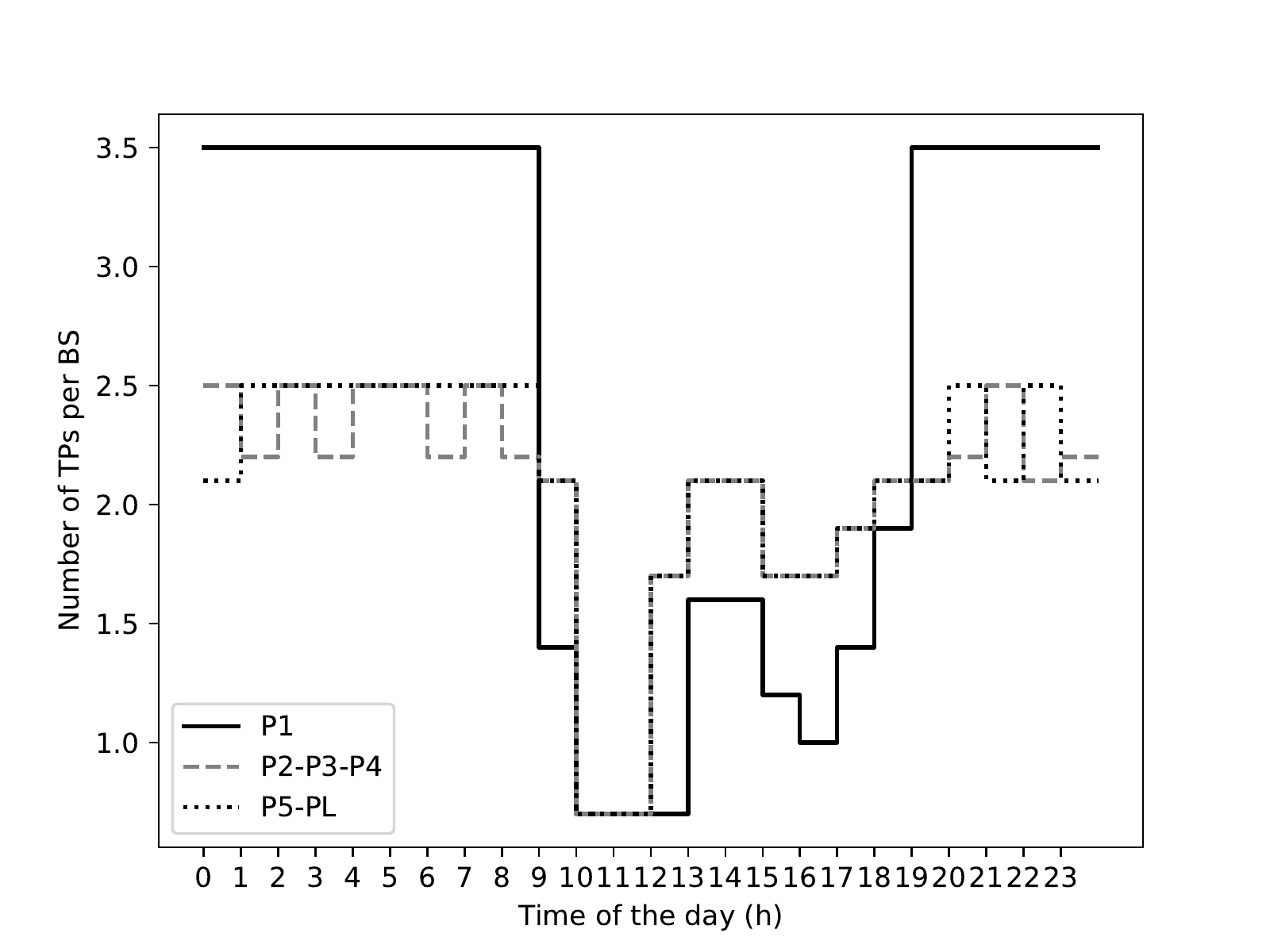}
    \caption{$\sigma$ Profiles}
    \label{fig:std}
\end{figure}
           
\subsection{User Assignments}
\label{sec:tps-assign}
In this section, we study the assignment of users to the base stations
for the small network composed 12 test points and 4 base stations. The goal is to see
how balanced in average is that assignation for each one of the six
different problems.  We are focus on certain periods
of the day where the network dynamics are more obvious.
\subsubsection{Measurements}
\label{sec:measures}
In addition to the visual representation of the assignments from
Figure~\ref{fig:tps-assign}, we compute two measures to evaluate the
assignation of the test points  for every time slots: the dispersion and the
standard deviation of the assignment vector. With this
measurement, it is easier to see how balanced the network is
throughout the day for the every optimization problem.  

The dispersion is simply the difference between the maximum and minimum numbers of test points assigned to a single base station. Define the number of test points assigned to a given base station $j$ at time $t$ as the degree
\begin{equation}
\Delta_{j,t} = \sum_{i \in I} h_{i,j,t}  \label{eq:deg} \quad \forall j \in S, \ \forall t \in T. 
\end{equation}
At time $t$, the dispersion $D_t$ is 
\begin{equation}
D_{t} = \max_{j \in S}\Delta_{j,t} - \min_{j \in S}\Delta_{j,t}  \label{eq:disp} \quad \forall t \in T. 
\end{equation}
Next, we compute the standard deviation $\sigma_t$
\begin{align}
\sigma_{t} = \frac{1}{|S|}\sqrt{\sum_{j \in S} (\Delta_{j,t} - N^{tp}_{bs})^2}  \label{eq:std} \quad \forall t \in T,
\end{align}
where $|S|$ is the total number of base stations and $N^{tp}_{bs}$ the number of average test points connected per base station. This average value is equivalent to the ratio of
test points over installed base stations, which is an input parameter of our model described in section~\ref{sec:paramdef}. 

\subsubsection{Assignment results}
\label{sec:assign-res}
Figure~\ref{fig:tps-assign} shows the map of the network  the links
represent the 
assignment of a test point to a base station for the joint problem (PL). Four time
periods have been chosen to give an idea of the network dynamics:
10:00, 16:00, 17:00 and 18:00. Figure~\ref{fig:time-10} shows the
network when traffic is at its peak. Next, we can better see the
interaction between base stations~0 and~1 in
figures~\ref{fig:time-16},~\ref{fig:time-17}
and~\ref{fig:time-18} where a few test points are served alternatively by both
base stations in order to switch them off more often.  

We also present the dispersion in Figure~\ref{fig:disp} where it is shown the different profiles for each of the six problems. Note that problems P2, P3 and P4 and problems P5 and PL have the same measure of dispersion. The same goes for the $\sigma$ represented in Figure~\ref{fig:std}. Both of these figures are very similar and the same conclusions can be made with either one of these. 
In the peak period shown in~\ref{fig:time-10}, all of the base stations need to be activated. This peak period is translated into a smaller dispersion and standard deviation as shown at times 10:00 and 11:00 in figures~\ref{fig:disp} and~\ref{fig:std}. This is also true, yet less striking, in the second peak starting at 15:00 and ending at 17:00. On the other hand, the network is less balanced with a higher dispersion and $\sigma$ during night time because of the sleep mode of the base stations. In other words, with a lower traffic, some base stations are put to idle mode and will not have any test points connected which results in a greater dispersion and $\sigma$.  

At last, we focus on the different test point assignments for the six
problems. For the basic problem P1, since the network is not large, the assignment is therefore more randomized with a higher dispersion and $\sigma$ obtained in off-peak periods. For the other five problems, the maximum value of the two measurements is lower. The minimum stays the same because, in high traffic, the network is well balanced for every problem. P2-P3-P4 and P5-PL only differ during the off-peak before 9:00 and after 20:00. In these hours, the optimal solution of P5-PL has a  slightly greater dispersion and $\sigma$. 

Altogether, we see that the networks with both sleep mode and solar
energy are much more flexible and can  change the base
stations much more often than the other cases. This is of course the
root of the better efficiency of these networks as expected.
\section{Conclusion}
\label{sec:conclusion}
In this paper, we investigated the relationship between installing a solar harvesting system to power base station of a cellular network and the
energy management under varying demand. 
For this, we presented a solar installation planning model that takes into account the hourly dynamics of the cellular network. 
We challenged the belief that solar energy can be considered free and that should always be installed everywhere in the network.
This was done 
by explicitly modelling solar panels, batteries, inverters and charge controllers, as well as the cellular network demand and energy-management.
We found that the solar installation and the energy-management of the
base stations are so tightly interrelated that even the order in
which the  technologies are introduced can have an important impact on
network cost and network performance.
Finally, we show that installing solar equipment everywhere need not
be the best solution even when the unit cost of solar power is smaller
than that of the grid.

\bibliographystyle{plain}

\end{document}